\documentclass[]{article}  

\usepackage{graphics,epsfig}
\font\smallroman=cmr10 at 8pt

\title{\bf The stuff the world is made of: physics and reality\footnote{Published as: Aerts,
D., 1999, ``The stuff the world is made of: physics and reality", in {\it Einstein meets Magritte: an
interdisciplinary reflection}, eds. Aerts, D., Broekaert, J. and Mathijs, E., Kluwer Academic, Dordrecht.}}

\author{Diederik Aerts}

\begin{document}           
\date {} 

\maketitle

\centerline{Center Leo Apostel (CLEA), and}
\centerline {Foundations of the Exact Sciences (FUND),}
\centerline{Faculty of Science,}
\centerline{Brussels Free University,}
\centerline{Krijgskundestraat 33,1160 Brussels,}
\centerline{e-mail: diraerts@vub.ac.be}

\begin{abstract}
\noindent Taking into account the results that we have been obtained during the last decade in the foundations of quantum
mechanic we put forward a view on reality that we call the `creation discovery view'. In this view it is made explicit that
a measurement is an act of a macroscopic physical entity on a microphysical entity that entails the creation of new
elements of reality as well as the detection of existing elements of reality. Within this view most of the quantum
mechanical paradoxes are due to structural shortcomings of the standard quantum theory, which means that our analysis
agrees with the claim made in the Einstein Podolsky Rosen paper, namely that standards quantum mechanics is an incomplete
theory. This incompleteness is however not due to the absence of hidden variables but to the impossibility for standard
quantum mechanics to describe separated quantum entities. Nonlocality appears as a genuine property of nature in our view
and makes it necessary to reconsider the role of space in reality. Our proposal for a new interpretation for space makes it
possible to put forward an new hypothesis for why it has not been possible to unify quantum mechanics and relativity
theory.
\end{abstract}

\bigskip

\begin{quotation}

\noindent {\it Absolute space, in its own nature, without relation to anything external, remains always similar
and immovable. Absolute, true, and mathematical time, of itself, and from its own nature, flows equally without
relation to anything external.}

{\it Isaac Newton, 1642 - 1726.}
\end{quotation}

\begin{quotation}
\noindent {\it An intelligence that would know at a certain moment all the forces existing in nature and the
situations of the bodies that compose nature, and if it would be powerful enough to analyze all these data, would
be able to grasp in one formula the movements of the biggest bodies of the Universe as well as of the lightest
atom.}


{\it Simon Laplace, 1749 - 1827}
\end{quotation}

\begin{quotation}
\noindent {\it Because of the relativity of the concept of simultaneity, space and time melt together to a four
dimensional continuum.}

{\it Albert Einstein, 1897 - 1955}
\end{quotation}

\begin{quotation}
\noindent {\it Everything is still unclear to me, but my feeling is getting stronger everyday. I believe that in
the scheme that I am developing the particles will not move anymore on orbits, and we shall have to reconsider
fundamental classical concepts.}

{\it Werner Heisenberg, 1901 - 1976}

\end{quotation}

\noindent The word Physics comes from the Greek word `phusis', which means `that what comes into existence', and
itself is derived from the Greek verb `phuoo' which means `to create, to come into existence'.

In this paper we want to investigate what we can say about reality taking into account the latest insights from
physics. We shall see that our intuitive conception of reality is challenged by the two fundamental physical
theories of modern times, quantum mechanics and relativity theory. Instead of starting here with subtle
philosophical considerations - we shall have ample place for that later - we want to confront the reader
immediately with one of the more mysterious aspects of quantum reality, namely, non-locality.

\section{ Magic with neutrons.}

In this section we present an experiment on single quantum entities that illustrates, in our opinion, the problem
of non-locality as encountered in quantum mechanics in its most crucial form. It is an experiment in neutron
interferometry performed by Helmut Rauch and his collaborators. The preparation of the experiment was published
in [1], while the actual experiment, as presented here, was performed a year later and the results were published
in [2]. Rauch has also written a `review article' on the numerous neutron experiments that have since been
performed [3]. 

Helmut Rauch and his group had built their first neutron interferometer in 1976. To do this, starting from a
perfect monocrystalline silicium block, they had cut out a crystal in the shape shown in Figure 1, with three
parallel walls or lips of precisely the same thickness. In their experiments, they directed a neutron beam onto
one side of the crystal lips, and detected it on the other side. According to quantum mechanics the beam should
behave in a rather mysterious manner, and Rauch and his group wanted to verify if the predicted behavior was
correct. The beam was directed onto the crystal from the ``northwest" direction (see Figure 2). On the first lip,
the incident beam splits into two beams, which we shall call the northern and the southern beams; these then
travel on towards the second lip. 
\vskip 0.5 cm

\hskip 3.5 cm \includegraphics{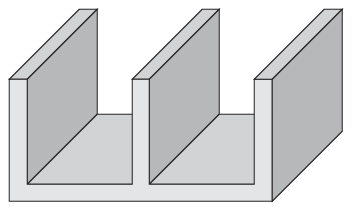}

\begin{quotation}
\noindent \baselineskip= 7pt \smallroman Fig. 1: The perfect silicium
crystal, as used by Helmut Rauch's group at the Laue-Langevin Institute in
Grenoble.
\end{quotation}
The northern beam undergoes refraction at the first lip, and travels on a
northeast course, while the southern beam continues in the prolongation of the incident beam. On the second lip,
the two beams again split, and of the four resultant beams, two will converge from north and south to cross on
the third lip.

\vskip 0.5 cm

\hskip 1.2 cm \includegraphics{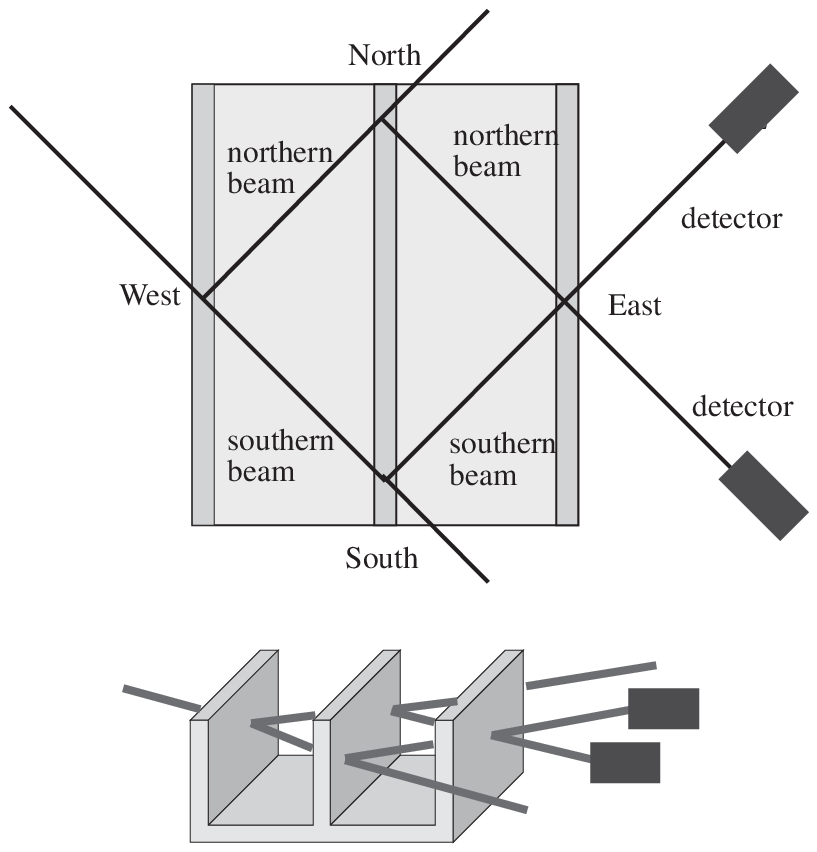}

\begin{quotation}
\noindent \baselineskip= 7pt \smallroman Fig. 2: Two representations of the Rauch experiment. The incident
beam comes in from northwest, and is split on the first lip into two beams: one is refracted to the northeast
(the northern beam), and the other (the southern beam) continues in the southeasterly direction. On the second
lip there is a further splitting giving rise to four beams, of which two, the northern and the southern beam,
cross on the third lip, and upon emerging from the crystal, are detected, and the neutrons counted.
\end{quotation}
Two detectors placed on their paths make it possible to count the neutrons as they emerge from the
crystal. Rauch's crystal is 7 cm long and 8 cm wide, so that the top view of Figure 2 is half of the real size.
The neutrons are emitted one at the time from a  reactor at an average speed of 2200 meters per second, which is
approximately 5000 miles per hour, and on average they are separated by a distance of 300 meters. This means that
there will never be more than one single neutron within the crystal. In point of fact, when a given neutron
passes through the crystal lips, the neutron that will follow has not yet been produced in the reactor.

In Rauch's experiments each of the neutrons has a ``coherence length" of one millionth of a centimeter. This means
that the region within which the neutrons exercises an action, or inversely, within which it can be acted upon,
is restricted to a cube of side one millionth of a centimeter. This is a very small volume indeed, and one of the
problems that we are confronted with is that we lose all intuitive feeling on such small scales. To understand
fully just how strange the results of Rauch's experiments are, let me scale the volume up to a size where we can
better visualize it. Let us therefore reconsider the Rauch experiments on a scale 25 million times larger.

To do this, first take the real crystal and place it on a map of Europe scaled down twenty five million times.
Then scale back up to get an imaginary super crystal covering a large area of Central Europe (Figure 3). The
neutrons will now seem to be coming in from over the Atlantic Ocean, penetrating the super-crystal in Paris. The
first lip, in which the neutron beam is split, lies over France and Great Britain. The northern beam flicks
north-east over Belgium, and penetrates the second lip somewhere between Denmark and Norway. The southern beam
passes over Bern, and attains the second lip in Trieste. In the second lip, the beams are are again split in two,
so that four beams emerge, of which two in the direction of Warsaw where they cross. The northern beam has passed
over Copenhagen, and the southern over Vienna. Upon emerging from the crystal, the neutrons fly on towards Saint
Petersburg or the Crimea, where they will be detected. We mentioned that in the real experiments the field of influence of the neutrons can be considered as
localized within a small cube of side one millionth of a centimeter. This becomes a cube of 25 centimeters on the
scale for which the crystal covers half of Europe.

The passage of the neutron beam through the crystal lips will probably have suggested the following picture in
most readers' minds: the neutrons as small projectiles, and the beam as a machine-gun fire of these projectiles.
Let us think through a  Rauch experiment assuming that the projectile analogy is correct. The machine-gun which
is firing the neutrons lies somewhere over the Atlantic Ocean and is aiming at Paris. Remember that there is
never more that a single neutron within the crystal at any given moment. This means that our machine-gun fires
very slowly, one neutron after the other at large time intervals. A given neutron will have been detected in
Saint Petersburg or in the Crimea long before the next neutron is fired. In our projectile analogy we can thus
consider individual trajectories for each neutron taken separately. A given neutron comes in above the Channel,
penetrates the crystal in Paris, then either continues through towards Vienna on the southern beam line, or is
deflected towards Copenhagen on the northern beam. In the second lip, the same thing happens again: either the
neutron passes through undeflected and leaves the crystal, or it is deflected, and flicks over Vienna or
Copenhagen in the direction of Warsaw where it reaches the third lip. Yet again the neutron can proceed
undeflected, and it will finally reach the detectors in Saint Petersburg or the Crimea.

\vskip 0.5 cm

\hskip 0.7 cm \includegraphics{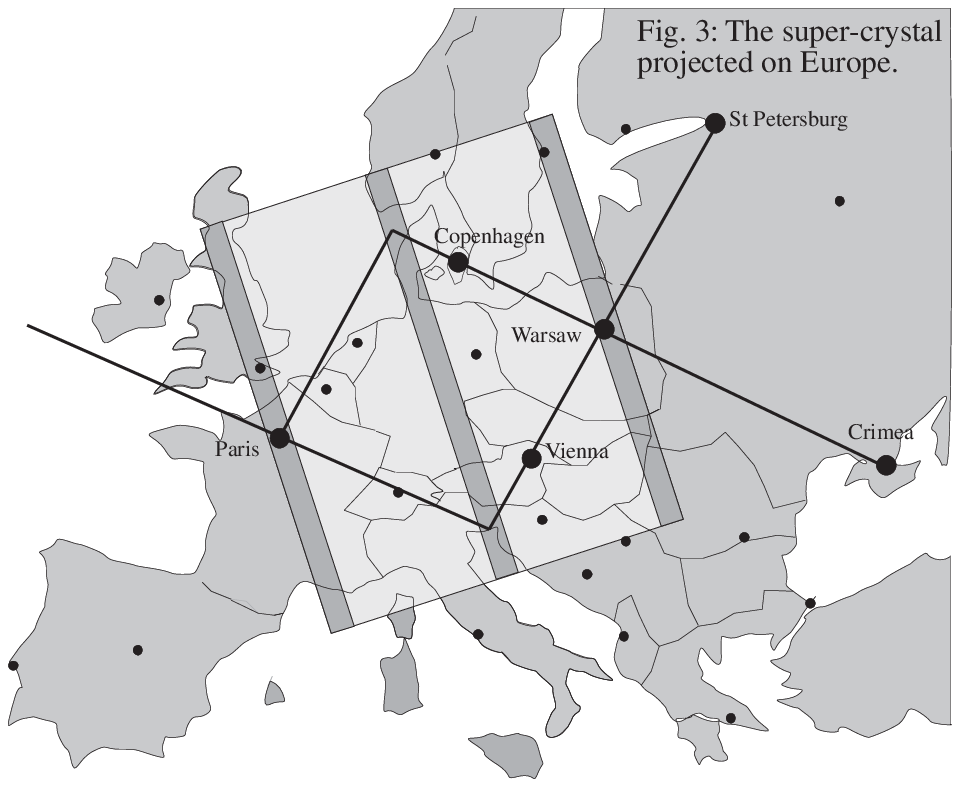}

\noindent 
If this machine-gun projectile analogy were correct, it would be difficult to imagine anything mysterious about
this experiment. But it is not correct. Further on, we shall give a complete quantum mechanical description of
what takes place, so that we shall be able to see step by step how the mystery arises. At present, let us just
consider what actually happens in Rauch's experiments, because that is our direct concern at present.

\begin{quotation}
\noindent {\it The experimental set-up is such that Rauch is able to act upon  each neutron as it crosses lip 2 of
the crystal, i.e. in our upscaled model, within a 25 centimeter  cube either  in Copenhagen, or in Vienna. More
precisely, Rauch can rotate the neutron, using experimental apparatus located in Copenhagen or Vienna, and which
has only a local effective range. The rotation of the neutron can be carried out from either of the two
experimental sites, Vienna or Copenhagen, and will be observed by one of the detectors, in  Saint Petersburg or
in the Crimea.}
\end{quotation}

\noindent From this it is clear that the neutron does not behave like a small projectile, for then it would pass
either through Copenhagen, and Rauch could not rotate it from Vienna, or it would pass through  Vienna, so that
he could not act on it from Copenhagen. The experiment establishes that it is truly possible to rotate the
neutron both from Copenhagen and from Vienna, without anything happening in the space between Vienna and
Copenhagen. No signal which could influence the neutron in any way is transmitted between Vienna and Copenhagen.

The apparatus which Rauch uses to rotate the neutron is a magnetic field localized in a small region in Vienna and
Copenhagen. There is no possibility whatsoever that the magnetic field used in Copenhagen to rotate the neutron
could have any action outside Copenhagen, let alone in Vienna; at least if we think of magnetic forces varying in
space. And there is no possibility that the neutron is partly in Copenhagen and partly in Vienna (whatever this
would mean), because, if we were to set up detectors there, what we would detect in Vienna or in Copenhagen would
always be either a complete neutron or no neutron. More specifically there is one chance out of two for the whole
neutron to be detected in Copenhagen and one chance out of two for it to be detected in Vienna. It is `as if' the
single neutron is present simultaneous in both places, in the small cube in Vienna and in the small cube in
Copenhagen, and that it can be acted upon from both these places as though it really and truly be there. An
object which is simultaneous present in two distant places, can such a thing possibly exist? Yet this is the
result predicted by the theory of quantum mechanics and obtained experimentally by Rauch. But quantum theory does
not tell us of how to understand this effect, and it is only recently that we are beginning to understand more of
it.

\section{Non locality and the concept of space.}

The Ptolomean system for our universe was not abandoned by reason of experimental errors, for it fitted very well
with all existing observations. To incorporate the descriptions of the known phenomena it only had to introduce
additional constructions, called {\it epicycles}, which gave rise to many complications but gave a good fit to
the experimental observations. But since the primary hypothesis {\it (a) the earth is the center of the universe}
and {\it (b) all celestial bodies move in circles around the earth} were felt to be absolutely essential, the
complications could be interpreted as being due to specific properties of the planets. Copernicus (and Greek
scientists long before him) dropped hypothesis {\it (a)}, substituting it by a new one {\it (c) the sun is the
center of the universe}. Clearly this new hypothesis gave rise to a model that is much simpler than the original
Ptolomaeus model. Until the theoretical findings of Kepler, using the refined experimental results of Brahe,
hypothesis {\it (b)}, the circle as the basic motion for the celestial objects, remained unaltered, and Kepler
was very unhappy when it became clear to him that it was a wrong hypothesis. Now that we know the motion of the
planets around the sun as a general solution of Newton's equations, the fact that these motions proceed along
ellipses does not bother us anymore. On the contrary, the elliptic orbits have become a part of a much greater
whole, Newtonian mechanics, which incorporate more beauty and symmetry than the original two axioms that were of
primary importance to Ptolomaeus.  

\par The change from Ptolomaeus to Copernicus is typical in the evolution of scientific theories. Usually one is
not conscious of the concepts that prevent scientific theories from evolving in a fruitful direction. We claim
that we have now a similar situation for quantum mechanics, and that the concept of quantum entity, and its
meaning, is at the heart of it. We believe that the pre-scientific preconception that has to be abandoned can be
compared to that of the earth being the center of the universe. It is a preconception that is due to the specific
nature of our human interaction with the rest of reality, and of the subjective perspective following from this
human interaction. We can only observe the universe from the earth, and this gave us the perspective that the
earth plays a central role. In an analogous way we can only observe the micro-world from our position in the
macro-world; this forces us to extend the concepts of the worldview constructed for this macro-world into the
worldview that we try to construct for the micro-world. That space-time is the global setting for reality is such
a hypothesis, and it leans only on our experience with the macroscopic material world.

The experiment with the neutrons is only one of the many experiments that have been carried out recently to
exhibit the quantum effect that has been called {\it non-locality}. We cannot go into all details in this paper
and refer the reader to [1, 2, 3, 4, 5, 6] for extensive analyses of Rauch's experiment. Meanwhile, more than two
decades later, experimentators play in the laboratory with quantum entities brought very explicitly in non-local
states . And in 1997, Nicolas Gisin - with whom the author of this article made his first steps in research as
young students at the university of Geneva - managed to produce a pair of non-local photons over a distance of 20
kilometers, using glass fibers of Swiss Telecom between two Swiss villages. All this shows that non- locality is
a genuine property of quantum entities.

It is our opinion that one cannot retain in quantum mechanics the hypothesis that at every moment every entity is
effectively present in space. The behaviour of quantum entities, not only in Rauch's experiment with neutrons but
also in many other experiments, shows us that this idea must be incorrect. Let us therefore explicitly introduce
the following hypothesis: 

\begin{quotation}
\noindent {\it We shall assume that quantum entities are not permanently present in space, and that, when a
quantum entity is detected in such a non-spatial state, it is `dragged' or `sucked up' into space by the
detection system}
\end{quotation}

\noindent In our everyday reality, each material entity has at every instant its place in space. In classical
mechanics, there are various ways of specifying position in 3-dimensional space. For a solid body, one can give
the position of its centre of gravity, and its orientation in a coordinate system with origin in the centre of
gravity. For a liquid or a gas, one will use continuum mechanics and a description in terms of fluid particles,
filling that part of space where the mass density of the liquid or the gas is different from zero. Waves too,
although often spread out, can be given a place in space. In classical mechanics, whatever the description used
and whichever entity is considered, it is in a well defined place. In the picture that we now want to propose for
quantum entities the situation is very different. We assume that the experiment in which the quantum entity is
detected contains a creation-element: this actually in part creates a place for the entity, at the moment when
the detection is carried out. More explicitly this means that, before the experiment, the quantum entity did not
necessarily have a place in space and that its place is created by the experiment itself. An analogous process
happens when the momentum (product of velocity times mass and intuitively thought of as the impact that an entity
has on another entity when colliding) of the quantum entity is measured. The quantum entity will not in general
have a momentum before the experiment carried out to measure it.  As often happens, everyday language helps us to
understand this change in meaning of the concept of space. One often considers reality as the setting in which
everything takes place. Events, when we do not yet consider them as entities, still `take place', which can be
considered to imply that they are not necessarily in space to start with.

At first sight it might seem that such a picture cannot satisfy those scientists who seek the intuitive support of
their imagination; but we shall see that this impression is erroneous, and that it comes from preconceived ideas
over what `being' really is.

This brings us thus to the central question: the nature of `being', or in other words, the nature of reality. 

It is perhaps now the moment to say that the results that we present in this article have been acquired over a
period of two decades. In a first period, mostly by myself, but certainly inspired by my experience as doctorate
student in the school of Constantin Piron in Geneva. And later, together with my young collaborators in our
research group FUND at the university of Brussels. The totality of our results together form a specific view, an
interpretation of quantum mechanics, that we have called the {\it creation-discovery view}. We refer to [4, 7, 8,
9, 10, 11, 12] and first give here a short description of this {\it creation-discovery view}.

Within the {\it creation-discovery view} it is taken for granted that during an act of measurement there always
exist two aspects, a {\it discovery} of a part of reality that was present independently of the act of
measurement, and a {\it creation} that adds new elements of reality to the process of measurement and to the
entity under investigation. When we put forward the {\it creation- discovery view} in this way, it does not seem
to contradict our intuition. Indeed creating part of reality during the act of measurement is certainly not
contra- intuitive. We are confronted with so many situations in our daily life where such creation aspects are
obviously present. To make clear what we mean let me give a very common example from our everyday life. Suppose
that an interviewer is questioning a person for an opinion poll. It is obvious that the act of interviewing
itself, the way the question is asked, the attitude of the interviewer, in short, each aspect of the context in
which the interview takes place, can in part create the answer of the person interviewed, depending on the type
of question that is asked. In this example of the interview for an opinion poll, the creative aspect is well
known and not mysterious at all. The {\it creation- discovery view} as applied to the interpretation of quantum
mechanics has however far reaching consequences that do contradict certain aspects of our intuition. More
precisely, and we come back now to the situation of Rauch's experiment,  it is the hypothesis that the whole of
reality can be contained within space that turns out to be at stake. Indeed, we can show that within {\it `the
creation-discovery view'} as applied to the micro-world, the creation aspect of a quantum measurement in the
detection of a quantum entity contains in part the creation of the {\it place} of the quantum entity itself. This
means that the place of this quantum entity {\it did not} exist before the entity was detected, and this place is
created during the process of detection. The same is true for the property `momentum' of a quantum entity. It is
partly created during the process of the measurement of this property, and did not exist before. As a 
consequence, a quantum entity in most of its states does not have a place - in technical jargon, we say that it
is {\it not localized} - and it does not have a momentum (or impact which is a property more easy to imagine for
us). We want to state clearly that the reason we have developed this {\it creation-discovery view} is not because
we just wanted to try it out for philosophical purposes. The reason is that we were compelled to formulate it, on
the one hand, due to the new and very subtle experiments on single quantum entities like the one of Helmut Rauch,
and on the other hand, as a result of new theoretical investigations, the details of which are however too
technical to be presented here.

Let us now analyse in which way we apply the {\it creation-discovery view} to describe the experiments like the
one of Helmut Rauch. Within the {\it creation- discovery view}, we propose that the mysterious aspects of the
Rauch experiments result from the fact that the neutron involved is `not present in space'. And that the two
experimental cubes, the one in Copenhagen and the one in Vienna, can be considered as windows through which we
can act on the neutron in its non-spatial state. The two cubes are openings which give us access to the reality
`out of space'.

We no longer visualize space as an all-embracing setting in which the whole play of reality takes place, but as a
structure that we, as human beings, have constructed, relying upon our everyday experience of the macroscopic
entities around us. We make a distinction between the following two properties: 1) Every entity can be detected
in space, and space is then one of the structures in which we, as human beings, come into contact with and create
a reality. 2) Every entity is present in space, and space is then the setting in which all of reality develops.
While the first property also applies to quantum entities, the second does not.

Our {\it creation-discovery view} introduces a new quality of reality for space. Space as an intermediate
structure in which encounters occur, rather than as an all embracing setting. Things make their place instead of
having a place. Yet again, language is clearer in terms of events than of entities: think of the expression:
`participate' in the making of an event. In our {\it creation-discovery view} we participate in the making of an
entity. We actually suspect that it is the failure of space as a global setting for reality which accounts for
the unsuccessful outcome up till now of every attempt to unify relativity theory, for which space is a
fundamental ingredient, with quantum mechanics. We shall turn to this question in a later section  of this paper.

\section {The epicycles of De Broglie and Bohm, waves and particles.}

\par The pictures that have been put forward in a last but hard struggle to fit quantum entities within the
space-time setting make use of two basic prototypes: particles and waves. The particle is identified by the fact
that upon detection it leaves a spot on the detection screen, while waves are to be recognized by their
characteristic interference patterns. Certain experiments with quantum entities give results which are
characteristic for particles, other experiments reveal the presence of waves. This is the reason why the concepts
of particles and waves are used to attempt to represent quantum entities. 

\smallskip
\noindent {\it (1) De Broglie and Bohm: particles and waves.}
\smallskip

\noindent
There exists a representation using waves and particles together, introduced by Louis de Broglie [13] in the early
years of quantum mechanics, and which after a long period of neglect, was rediscovered by David Bohm and Jean
Pierre Vigier [14] and which is still now the object of active study in different research centers. In this
representation, it is assumed that a quantum entity is at the same time always both a particle and a wave. The
particle has the properties of a small projectile, but is accompanied by a wave which is responsible for the
interference patterns. This representation of de Broglie and Bohm incorporates the observed quantum phenomena and
attempts to change as little as possible at the level of the underlying reality where these quantum entities
exist and interact. This reality is the ordinary three-dimensional Euclidean space; the quantum entity is
considered to be both a wave and a particle, existing, moving and changing in this space. The specific quantum
effects are accounted for by a quantum potential which is effective in this three dimensional Euclidean space,
and which brings about the quantum non-local effects. The quantum potential is the entity that carries most of
the strange quantum behavior. The quantum probabilities appear in the de Broglie-Bohm picture as ordinary
classical probabilities, resulting from a lack of knowledge about the position of the point particle associated
with the quantum entity. This is exactly as for the probabilities in a classical statistical theory, and is due
to a lack of knowledge about the micro-states of the atoms and molecules of the substance considered. The de
Broglie-Bohm picture is thus a hidden variable theory. The variables describing the state of the point particle
are the hidden variables, and the lack of knowledge about these hidden variables is at the origin of the
probabilistic description.

\par There is however a serious problem with the de Broglie-Bohm theory when one attempts to describe more than
one quantum entity. Indeed, for the example of two quantum entities, the wave corresponding to the composite
entity consisting of the two quantum entities is a wave in a six dimensional configuration space, and not in the
three-dimensional Euclidean space, and the quantum potential acts in this six-dimensional configuration space and
not in the three-dimensional Euclidean space. Moreover, when the composite entity is in a so called {\it
`non-product state'}, this wave in the six-dimensional configuration space cannot be written as the product of
two waves in the three-dimensional space (hence the reason for naming these states ``non-product states"). It is
these non-product states that give rise to the typical quantum mechanical Einstein-Podolsky-Rosen-like
correlations between the two sub-entities. The existence of these correlations has meanwhile been experimentally
verified by different experiments, so that the reality of the non-product states, and consequently the
impossibility to define the de Broglie-Bohm theory in three-dimensional space, is firmly established. This
important conceptual failure of the de Broglie-Bohm theory is certainly also one of the main reasons that Bohm
himself considered the theory as being a preliminary version of yet another theory to come [15].  

\smallskip
\noindent {\it (2) Bohr: the Copenhagen interpretation.}
\smallskip

\noindent
The usual representation of quantum entities makes use either of a wave or of a particle, and although it is now
associated with the Copenhagen school, it was present in quantum mechanics from the very start. In this picture
it is considered that the quantum entity can behave in two ways, either like a particle or like a wave, and that
the choice between the two types of behavior is determined by the nature of the observation being made. If the
measurement one is making consists in detecting the quantum entity, then it will behave like a particle and leave
a spot on the detection screen, just as a small projectile would. But if one chooses an interferometric
experiment, then the quantum entity will behave like a wave, and give rise to the typical interference pattern
characteristic for waves. When referring to this picture one usually speaks of Bohr's complementarity principle,
thereby stressing the dual structure assumed for the quantum entity. This aspect of the Copenhagen interpretation
has profound consequences for the general nature of reality. The complementarity principle introduces the
necessity of a far reaching subjective interpretation for quantum theory. If the nature of the behavior of a
quantum entity (wave or particle) depends on the choice of the experiment that one decides to perform, then the
nature of reality as a whole depends explicitly on the act of observation of this reality. As a consequence it
makes no sense to speak about a reality which exists independently of the observer. 

\par This dramatic aspect of the Copenhagen interpretation is best illustrated by the delayed-choice experiments
proposed by John Archibald Wheeler, where the experimental choice made at one moment can modify the past.
Wheeler's reasoning is based on an experimental apparatus as shown in Figure 4, where a source emits extremely
low intensity photons, one at a time, with a long time interval between one photon and the next. The light beam
is incident on a semitransparant mirror $A$ and divides into two beams, a northern beam $n$, which is again
reflected by the totally reflecting mirror $N$ and sent towards the photomultiplier $D_1$, and a southern beam
$s$, which is reflected by the totally reflecting mirror $S$, and sent towards the photomultiplier $D_2$. We know
that the outcome of the experiment will be that every photon will be detected either by $D_1$ or by $D_2$.

\vskip 0.5 cm

\hskip 2.2 cm \includegraphics{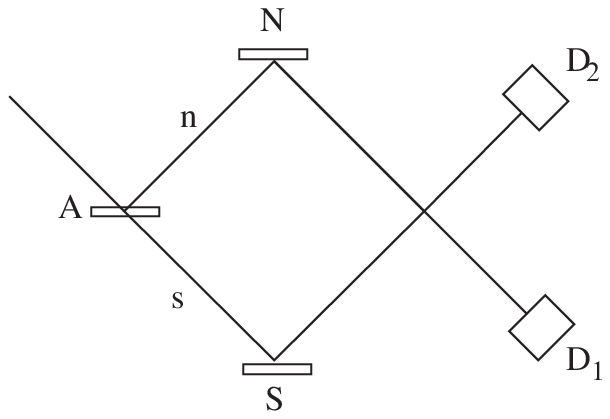}

\begin{quotation}
\noindent \baselineskip= 7pt \smallroman Fig. 4 : The delayed-choice
experimental setup as proposed by John Archibald Wheeler. A source emits extremely low intensity photons that are
incident on a semitransparant mirror $\scriptstyle A$. The beam divides into two, a northern beam $\scriptstyle
n$, which is again reflected by the totally reflecting mirror $\scriptstyle N$ and sent towards the
photomultiplier $\scriptstyle D_1$, and a southern beam $\scriptstyle s$, which is reflected by the totally
reflecting mirror $\scriptstyle S$, and sent towards the photomultiplier $\scriptstyle D_2$.
\end{quotation}
Following the Copenhagen complementarity interpretation, this experimental situation forces a photon to behave
like a particle, that will be detected either in the northern detector $D_2$ or in the southern detector $D_1$.
It is quite easy to introduce an additional element in the experimental setup, that according to the Copenhagen
interpretation will make the photons behave like a wave.

\vskip 0.5 cm

\hskip 2.2 cm \includegraphics{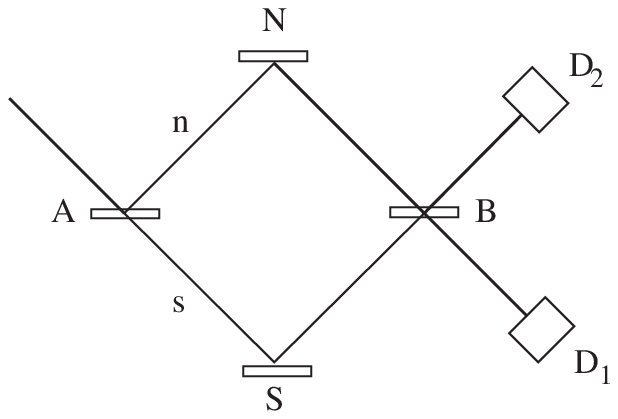}

\begin{quotation}
\noindent \baselineskip= 7pt \smallroman Fig. 5 : The delayed-choice
experimental setup as proposed by John Archibald Wheeler, where a second semitransparent mirror is introduced.
Following the Copenhagen interpretation, in this experimental situation the photons will behave like a wave.
\end{quotation}
Wheeler proposes the following: we introduce a second
semitransparent mirror $B$ as shown on Figure 5, and the thickness of $B$ is calculated as a function of the
wavelength of the light, such that the superposition of the northern beam and the southern beam generates a wave
of zero intensity. 

In this experimental setup nothing will be detected in $D_2$ and all the light goes to $D_1$, and the photons of
the beam are forced into a total wave behavior. Indeed, each photon interferes with itself in region $B$ such
that it is detected with certainty in $D_1$. So, we have two experimental setups, the one shown in Figure 1 and
the one shown in Figure 2, that only differ by the insertion of a semitransparent mirror $B$. Wheeler proposes
the semitransparent mirror $B$ to be inserted or excluded at the last moment, when the photon has already left
the source and interacted with the mirror $N$. Following the Copenhagen interpretation and this experimental
proposal of Wheeler, the wave behavior or particle behavior of a quantum entity in the past, could be decided
upon by an experimental choice that is made in the present. We are dealing here with an inversion of the
cause-effect relationship, that gives rise to a total upset of the temporal order of phenomena.

\par To indicate more drastically the profound subjective nature of the worldview that follows from a consistent
application of the Copenhagen interpretation, Wheeler proposes an astronomical version of his delayed-choice
experiment. He considers the observation on earth of the light coming from a distant star. The light reaches the
earth by two paths due to the presence of a gravitational lens, formed by a very massive galaxy between the earth
and the distant star. Wheeler observes that one may apply the scheme of Figure 1 and 2, where instead of the
semitransparent mirror $A$ there is now the gravitational lens. The distant star may be billions of light years
away, and by the insertion or not of the semitransparent mirror, we can force the next photon that arrives to
have traveled towards the earth in the form of a wave or of a particle. This means according to Wheeler that we
can influence the past even on time scales comparable to the age of the universe. 

\par Not all physicist who believe in the correctness of the Copenhagen interpretation go as far as Wheeler
proposes. The general conclusion of Wheeler's example remains however valid. The Copenhagen interpretation makes
it quite impossible to avoid the introduction of an essential effect on the nature and behavior of the quantum
entity due to the choice of the type of measurement that is performed on it. The determination of the nature and
the behavior of a quantum entity independently of the specification of the measurement that one is going to carry
out is considered to be impossible in the Copenhagen interpretation.

\smallskip
\noindent {\it (1) The creation-discovery view: quantum entities and space.}
\smallskip

\noindent
Let us explain now in which way the {\it creation- discovery view} that we want to bring forward is different
from both of the above mentioned interpretations, the de Broglie Bohm interpretation and the Copenhagen
interpretation. It is a realistic interpretation of quantum theory, in the sense that it considers the quantum
entity as existing in the outside world, independently of us observing it, and with an existence and behavior
that is also independent of the kind of observation to be made. In this sense it is strictly different from the
Copenhagen interpretation, where the mere concept of quantum entity existing independently of the measurement
process is declared to be meaningless. The creation-discovery view is however not like the de Broglie-Bohm
theory, where it is attempted to picture quantum entities as point particles moving and changing in our
three-dimensional Euclidean space, and where detection is considered just to be an observation that does not
change the state of the quantum entity. In the creation-discovery view it is taken for granted that measurements,
in general, {\it do change the state of the entity} under consideration. In this way the view incorporates two
aspects, an aspect of `discovery' referring to the properties that the entity already had before the measurement
started (this aspect is independent of the measurement being made), and an aspect of `creation', referring to the
new properties that are created during the act of measurement (this aspect depends on the measurement being made).

\section {The quantum machine: a general operational formalism providing a closer approach to the mystery.}

\noindent	The fact that it took so long to come to the kind of view that we propose, is largely due to the way in
which quantum mechanics arose as a physical theory. Indeed, the development of quantum mechanics proceeded in a
rather haphazard manner, with the introduction of many ill-defined and poorly understood new concepts. 

During its first years (1890-1925, Max Planck, Albert Einstein, Louis de Broglie, Hendrik Lorentz, Niels Bohr,
Arnold Sommerfeld, and Hendrik Kramers), quantum mechanics (commonly referred to as the {\it `old quantum
theory'}), did not even possess a coherent mathematical basis. In 1925 Werner Heisenberg [16] and Erwin
Schr\"odinger [17] produced the first two versions of the new quantum mechanics, which then were unified by Paul
Dirac [18] and John Von Neumann [19] to form what is now known as the orthodox version of quantum mechanics. The
mathematical formalism was elaborate and sophisticated, but the significance of the basic concepts remained quite
vague and unclear. The predictive success of the theory was however so remarkable that it immediately was
accepted as constituting a fundamental contribution to physics. However, the problems surrounding its conceptual
basis led to a broad and prolonged debate in which all the leading physicists of the time participated (Einstein,
Bohr, Heisenberg, Schr\"odinger, Pauli, Dirac, Von Neumann, etc.)

\par The Von Neumann theory constitutes the standard mathematical model of quantum mechanics [19]. We give now a
short description of this standard model. Those readers who are not acquainted with the jargon, are advised just
to skip the next paragraph, and proceed.  

\begin{quotation}
\noindent {\it Standard quantum mechanics:} the state  of a quantum entity is described by a unit vector in a
separable complex Hilbert space; an experiment is described by a self-adjoint operator on this Hilbert space,
with as eigenvalues the possible results of the experiment. As the result of an experiment, a state will be
transformed into the eigenstate of the self-adjoint operator corresponding to a certain experimental result, with
a probability given by the square of the scalar product of the state vector and of the eigenstate unit vector. It
follows that, if the state of the quantum entity is not an eigenstate of an operator associated with a given
experiment, then the experiment can yield any possible result, with a probability determined by the scalar
product of the state and eigenstate vectors as indicated above. The dynamical evolution of the state of a quantum
entity is determined by the Schr\"odinger equation. \end{quotation}

\noindent The orthodox quantum mechanics of Von Neumann is still dominant in the classroom, although a number of
variant formalisms have since been developed with the aim of clarifying the basic conceptual shortcomings of the
orthodox theory. In the sixties and seventies, new formalisms were being investigated by many research groups. In
Geneva, the school of Josef Maria Jauch was developing an axiomatic formulation of quantum mechanics [20], and
Constantin Piron gave the proof of a fundamental representation theorem for the axiomatic structure [21]. Gunther
Ludwig's group in Marburg [22] developed the convex ensemble theory, and in Amherst, Massachusetts, the group of
Charles Randall and David Foulis [23, 24] was elaborating an operational approach. Peter Mittelstaedt and his
group in Cologne studied the logical aspects of the quantum formalism [25], while other workers (Jordan, Segal,
Mackey, Varadarajan, Emch) [26, 27] focused their attention on the algebraic structures, and Richard Feynman
developed the path integral formulation [28]. There appeared also theories of phase-space quantization, of
geometric quantization and quantization by transformation of algebras.

\par These different formalisms all contained attempts to clarify the conceptual labyrinth of the orthodox
theory, but none succeeded in resolving the fundamental difficulties. This was because they all followed the same
methodology: first develop a mathematical structure, then pass to its physical interpretation. This is still the
procedure followed in the most recent and authoritative theoretical developments in particle physics and
unification theory, such as quantumchromodynamics and string theory. But from 1980 on, within the group of
physicists involved in the study of quantum structures, there arose a growing feeling that a change of
methodology was indispensable, that one should start from the physics of the problem, and only proceed to the
construction of a theory after having clearly identified all basic concepts. Very fortunately, this change in
attitude to theory co\"{\i}ncided with the appearance of an abundance of new experimental results concerning many
subtle aspects of microphysics, which previously could only have been conjectured upon. We here have in mind the
experiments in neutron interferometry, in quantum optics, on isolated atoms, etc. The new insights as to the
nature of physical reality, resulting in part from the new experimental data and in part from the new
methodological approach, have made it possible to clarify some of the old quantum paradoxes and thereby to open
the way to a reformulation of quantum mechanics on an adequate physical basis.

 In Brussels we have now decided to work explicitly along this new methodological approach, starting from the
physics of the problem, and only proceeding to the construction of a theory after having clearly identified all
basic concepts [11, 29, 30, 31]. We however want to state clearly the following. One could get the impression
that such an approach, starting from the physics of the situation and then introducing the mathematics, solves
the old problem of operationality. Indeed, such a theory is by definition operational, since the basic
mathematical concepts are linked to well known `operations' and `situations' in the physical world.
Philosophically speaking however we do not believe that in this way we shall be able to reduce quantum mechanics
to a purely operational theory. We do not believe this because we are convinced of the fact that the micro-world
contains fundamentally new aspects of reality that cannot be reduced operationally to aspects of reality that we
take from the macroscopic world that surrounds us. But, we do think that an operational approach has to be pushed
to the limit as far as it can, because in this way we shall be able to come closer to these new strange aspects
of reality of the microworld.

We shall now describe the basic steps of our approach, illustrating it by means of the very simple example of a
{\it quantum machine} [7, 8, 9, 10, 12], which we shall here use to explain that part of quantum mechanics that
can at present be understood. 

\smallskip
\noindent {\it (1) The ontological basis: the concept of entity and its states.}
\smallskip

\noindent 
An entity $S$ is in all generality described by the collection $\Sigma$ of its possible states. A state $p$, at
the instant $t$, represents the physical reality of the entity $S$ at the time $t$. It represents what the entity
{\it `is'} at the time t. We use the concept of the state $p$ in the following way: 

\begin{quotation}
\noindent {\it At each instant of time $t$ an entity $S$ is in a specific state $p$, that represents the reality
of the entity at the time $t$.}
\end{quotation} 

\noindent  We remark that no mathematical structure is a priori assigned to this collection of states, contrary to
what is done in quantum mechanics (a Hilbert space structure) or in classical mechanics (a phase space
structure). 

The quantum machine (denoted $qm$ in the following) that we want to introduce - to illustrate the concepts that
are defined in a more general way - consists of a physical entity $S_{qm}$ constituted by a point particle $P$
that can move on the surface of a sphere, denoted $surf$, with center $O$ and radius $1$. The unit-vector $v$
giving the location of the particle on $surf$ represents the state $p_v$ of the particle (see Fig. 6,a). Hence
the collection of all possible states of the entity $S_{qm}$ that we consider is given by $\Sigma_{qm} = \{p_v\
\vert\ v \in surf\}$.

\smallskip
\noindent {\it (2) Operational foundation: experiments and outcomes.}
\smallskip

\noindent We acquire knowledge about the reality of the entity by performing experiments. In this way to each
entity $S$ and its set of states $\Sigma$ there corresponds a collection of relevant experiments - we shall
denote this collection by ${\cal E}$ - that can be carried out on the entity $S$. For an experiment $e \in {\cal
E}$ we denote its outcome set by $O(e)$ and each outcome by $x(e)_i$, hence $O(e) = \{x(e)_i \vert i \in I \}$. 

\vskip 0.5 cm

\hskip 2.2 cm \includegraphics{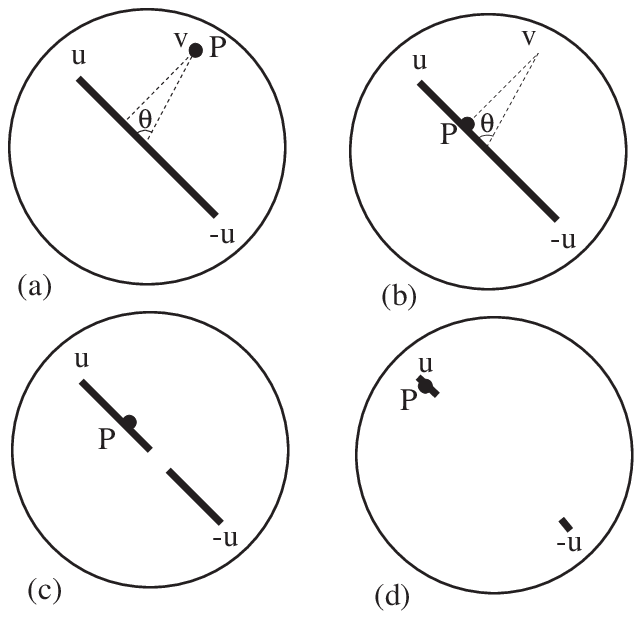}

\begin{quotation}
\noindent \baselineskip= 7pt \smallroman Fig. 6 : A representation of the quantum machine. In (a) the
physical entity $\scriptstyle P$ is in state $\scriptstyle p_v$ in the point $\scriptstyle
v$, and the elastic corresponding to the experiment $\scriptstyle e_{u}$ is installed between
the two diametrically opposed points $\scriptstyle u$ and  $\scriptstyle -u$. In (b) the
particle $\scriptstyle P$ falls orthogonally onto the elastic  and sticks to it. In (c) the
elastic breaks and the particle $\scriptstyle P$ is pulled  towards the point $\scriptstyle u$,
such that (d) it arrives at the point  $\scriptstyle u$, and the experiment $\scriptstyle e_u$
gets the outcome $\scriptstyle  o^u_1$.
\end{quotation}
Again, no a priori mathematical structure is imposed upon $\cal E$.

\begin{quotation}
\noindent {\it For an entity $S$ in a state $p$ an experiment $e$ can be performed and one of the outcomes
$o^e_i, i \in I$ will occur.}
\end{quotation}

\noindent For our quantum machine we introduce the following experiments. For each point $u \in surf$,
we introduce the experiment $e_u$. We consider the diametrically opposite point $-u$, and install an elastic band
of length 2, such that it is fixed with one of its end-points in $u$ and the other end-point in $-u$. Once the
elastic is installed, the particle $P$ falls from its original place $v$ orthogonally onto the elastic, and
sticks to it (Fig 6,b). The elastic then breaks and the particle $P$, attached to one of the two pieces of the
elastic (Fig 6,c), moves to one of the two end-points $u$ or $-u$  (Fig 6,d). Depending on whether the particle
$P$ arrives in $u$ (as in  Fig 6) or in $-u$, we give the outcome $x^u_1$ or $x^u_2$ to $e_u$. Hence for the
quantum machine we have ${\cal E}_{qm} = \{e_u\ \vert\ u \in surf\}$.

\smallskip
\noindent {\it (3) Change of state resulting from an experiment.}
\smallskip

\noindent If an experiment $e$ is performed on an entity $S$ in state $p$, and an outcome $x(e)_i$ occurs, this
state $p$ will in general be changed into one of the states $q_i, i \in I$ after the experiment.

\begin{quotation}
\noindent {\it For an entity $S$ in a state $p$ and an experiment $e$ with outcomes $x(e)_i$, the state $p$ is
changed into one of the states  $q_i, i \in I$ by the performance of the experiment $e$.} \end{quotation}

\noindent For the quantum machine the state $p_v$ is changed by the experiment $e_u$ into one of the two
states $p_u$ or $p_{-u}$.

\vskip 0.5 cm

\hskip 4 cm \includegraphics{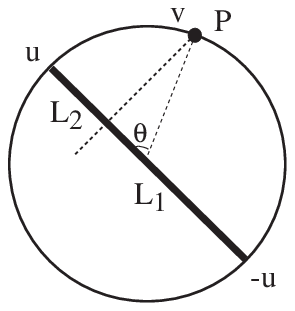}

\begin{quotation}
\noindent \baselineskip= 7pt \smallroman Fig. 7 : A representation of the experimental process in the
plane where it takes place. The elastic of length 2, corresponding to the experiment
$\scriptstyle e_u$, is installed between $\scriptstyle u$ and $\scriptstyle -u$.
The probability, $\scriptstyle P(x^u_1, p_v)$, that the particle  $\scriptstyle P$ ends up
in point $\scriptstyle u$ is given by the length of the piece of elastic $\scriptstyle
L_1$ divided by the total length of the elastic. The probability, $\scriptstyle P(x_2^u,
p_v)$, that the particle $\scriptstyle P$ ends up in point $\scriptstyle -u$ is given by the
length of the piece of elastic $\scriptstyle L_2$ divided by the total length of the
elastic.
\end{quotation}
\smallskip
\noindent {\it (4) Probability.}
\smallskip

\noindent
 For a given entity $S$ in a state $p$ and an experiment $e$ performed on this entity, each outcome $x(e)_i$ will
occur with a certain probability $P(x(e)_i, p)$, where this probability is the limit of the relative frequency of
repeated experiments. Hence we also have $\Sigma_i P(x(e)_i,p) = 1$.

\begin{quotation}
\noindent {\it For an entity $S$ in a state $p$ and an experiment $e$ with outcomes $\{x(e)_i \vert i \in I \}$,
there is a probability $P(x(e)_i,p)$ that the outcome $x(e)_i$ will occur and $\Sigma_i P(x(e)_i,p) = 1$.}
\end{quotation}

\noindent For the quantum machine we make the hypothesis that the elastic band breaks uniformly, which means that
the probability that a particle in state $p_v$, arrives in $u$, is given  by the length of $L_1$ (which is $1+
cos\theta$) divided by the total length of the elastic (which is 2). The probability that a particle in state
$p_v$ arrives in $-u$ is given by the length of $L_2$ (which is $1-cos\theta$) divided by the total length of the
elastic. If we  denote these probabilities respectively by $P(x^u_1, p_v)$ and $P(x^u_2, p_v)$ we have:  

\begin{eqnarray}
P(x^u_1, p_v) &=& {{1+cos\theta}\over 2} = cos^2{\theta\over 2} \\
P(x^u_2, p_v) &=& {{1-cos\theta}\over 2} = sin^2{\theta\over 2}  
\end{eqnarray}

\par \noindent In Figure 7 we represent the experimental process connected to $e_u$ in the plane where it takes
place, and  we can easily calculate the probabilities corresponding to the two possible  outcomes. In order to do
so we remark that the particle $P$ arrives in $u$ when the elastic  breaks in a point of the interval $L_1$, and
arrives in $-u$ when it breaks in a point of the  interval $L_2$ (see Fig. 7). 

\smallskip
\noindent
We have remarked already that in our approach we do not demand a priori any specific structure for the set of
states, for the set of experiments or for the probability model. This is one of the new and strong aspects of the
approach. One can question whether the structure that can be derived for such a situation is not too general to
be of any value. The method that we shall follow is however the following: for certain specific entities we shall
demand extra conditions to be fulfilled, but these conditions will also come from the physics of the situation
and will characterise exactly these specific entities. We refer the reader to [32] for a very detailed
outline of our operational and realistic approach.  

\smallskip
\noindent {\it (6) The quantum machine is a quantum entity.}
\smallskip

\noindent
We can easily show that the quantum machine is an entity the description of which is isomorphic to the quantum
description of the spin of a spin 1/2 particle. Hence, speaking in the quantum jargon, the quantum machine is a
model for the spin of a spin 1/2 quantum particle. This means that we can describe this macroscopic machine using
the ordinary quantum formalism with a two-dimensional complex Hilbert space as the carrier for the set of states
of the  entity.

The quantum machine as a model for an arbitrary quantum system described by a two dimensional Hilbert space was
presented in [7, 8, 9]. It is now possible to prove that for any arbitrary quantum entity one can construct a
model like that of the quantum machine [11, 33, 34, 35]. The explanation of the quantum structure that is given
in the quantum machine can thus also be used for general quantum entities. We have called this explanation the
`hidden measurement approach', hidden measurements referring to the fact that for a real measurement there is a
`lack of knowledge' about the measurement process in this approach. For the quantum machine, for example, this
lack of knowledge is the lack of knowledge about where the elastic will break during a measurement process.

This `physical' formalism has already led to a number of concrete and far reaching results, some of which we shall
explain in the following. The most important achievement however, in my opinion, consists in an explanation of
the structure of quantum mechanics, and in identifying the reason why it appears in a natural way in nature.

\section {What are quantum structures and why do they appear in nature?}

Already in the early development of quantum mechanics it was realized that the structure of quantum theory is very
different from the structure of the existing classical theories. This structural difference has been expressed
and studied in different mathematical categories, and we mention here some of the most important ones: 

\smallskip
\noindent
{\it (1) the structure of the collection of experimental propositions:}
\smallskip

\noindent 
If one considers the collection of properties (experimental propositions) of a physical entity, then it has the
structure of a Boolean lattice for the case of a classical entity, while it is non-Boolean for the case of a
quantum entity [20, 21, 36]

\smallskip
\noindent
{\it (2) the structure of the probability model:}
\smallskip

\noindent 
The axioms formulated by Kolmogorov in 1933 relate to the classical probability calculus as introduced for the
first time by Simon Laplace. Quantum probabilities do not satisfy these axioms. John Von Neumann was the first to
prove a ``no go" theorem for hidden variable theories [19]. Many further developments were however required 
before it was definitely proved that it is impossible to reproduce quantum probabilities from a hidden variable
theory. And quite definitely the structure of the quantum probability model is not Kolmogorovian [7, 8, 9, 23, 24,
37, 38, 39].

\smallskip
\noindent
{\it (3) the structure of the collection of observables:}
\smallskip

\noindent 
If the collection of observables is considered, a classical entity gives rise to a commutative algebra, while a
quantum entity does not [26, 27, 40, 41].

\smallskip
\noindent 
The presence of these deep structural differences between  classical theories and quantum theory has contributed
strongly to the belief that classical theories describe the ordinary `understandable' part of reality, while
quantum theory confronts us with a part of reality (the micro-world) that escapes our understanding. This is why
the strong paradigm that {\it quantum mechanics cannot be understood} is still in vigour. The example of our
macroscopic machine with a quantum structure challenges this paradigm, because obviously the functioning of this
machine can be understood. We now want to show that all the main aspects of the quantum structures can indeed be
explained in this way and we shall identify the reason why they appear in nature. We shall focus here on the
explanation in the category of the probability models, and refer to [11, 42, 43, 44, 45, 46, 47] for an analysis
pertinent to other categories.

The original development of probability theory aimed at a formalization of the description of the probabilities
that appear as the consequence of {\it a lack of knowledge}. The probability structure appearing in situations of
lack of knowledge was axiomatized by Kolmogorov and such a probability model is now called Kolmogorovian. Since
the quantum probability model is not Kolmogorovian, it has now generally been accepted that the quantum
probabilities are {\it not} associated with a {\it lack of knowledge}. Sometimes this conclusion is formulated by
stating that the quantum probabilities are {\it ontological} probabilities, as if they were present in reality
itself. In the approach that we follow in Brussels, and which we have named the {\it hidden measurement
approach},  we show that the quantum probabilities can also be explained as being due to a {\it lack of
knowledge}, and we prove that what distinguishes quantum probabilities from classical Kolmogorovian probabilities
is the {\it nature of this lack of knowledge}. Let us go back to the quantum machine to illustrate  what we mean.

If we consider again our quantum machine (Fig. 6 and Fig. 7), and look for the origin of the probabilities as they
appear in this example, we can remark that the probability is entirely due to a {\it lack of knowledge} about the
measurement process. Namely the lack of knowledge of where exactly the elastic breaks during a measurement. More
specifically, we can identify two main aspects of the experiment $e_u$ as it appears in the quantum machine. 

\begin{quotation}
\noindent (1) The experiment $e_u$ effects a real change on the state $p_v$
of the entity $S$. Indeed, the state $p_v$ changes into one of the states $p_u$ or $p_{-u}$
by the experiment $e_u$.
\end{quotation}
\begin{quotation}
\noindent (2) The probabilities appearing are due to a {\it lack of knowledge} about a
deeper reality of the individual measurement process itself, namely where the elastic breaks.
\end{quotation}

\noindent These two effects give rise to quantum-like structures, and the lack of knowledge about the deeper
reality of the individual measurement process comes from `hidden  measurements' that operate deterministically in
this deeper reality [7, 8, 9, 12, 48, 49, 50, 51, 52]; and that is the origin of the name that we gave to this
approach. 

One might think that our `hidden-measurement' approach is in fact a `hidden-variable' theory. In a certain sense
this is true. If our explanation for the quantum structures is the correct one, quantum mechanics is compatible
with a deterministic universe on the deepest level. There is no need to introduce the idea of an ontological
probability. Why then the generally held conviction that hidden variable theories cannot be used for quantum
mechanics? The reason is that those physicists who are interested in trying out hidden variable theories, are not
at all interested in the kind of theory that we propose here. They want the hidden variables to be hidden
variables of the state of the entity under study, so that the probability is associated to a lack of knowledge
about the deeper reality of this entity; as we have mentioned already this gives rise to a Kolmogorovian
probability theory. This kind of hidden variables relating to states is indeed impossible for quantum mechanics
for structural reasons, with exception of course in the de Broglie-Bohm theory: there, in addition to the hidden
state variables, a new spooky entity of `quantum potential' is introduced in order to express the action of the
measurement as a change in the hidden state variables; and as we have already remarked, the description of more
than one entity causes deep problems. 

If one wants to interpret our hidden measurements as hidden variables, then they are hidden variables of the
measuring apparatus and not of the entity under study. In this sense they are highly contextual, since each
experiment introduces a different set of hidden  variables. They differ from the variables of a classical hidden
variable theory, because they do not provide an `additional deeper' description of the reality of the physical
entity. Their presence, as variables of the experimental apparatus, has a well defined philosophical meaning, and
expresses the fact that we, human beings, want to construct a model of reality independent of our experience of
this reality. The reason is that we look for `properties' or `relations between properties', and these are
defined by our ability to make predictions independent of our experience. We want to model the structure of the
world, independently of our observing and experimenting with this world. Since we do not control these variables
in the experimental apparatus, we do not allow them in our model of reality, and the probability introduced by
them cannot be eliminated from a predictive theoretical model. In the macroscopic world, because of the
availability of many experiments with negligible fluctuations, we find an `almost' deterministic model.  \par We
must now try to understand the consequences of our explanation of the quantum structure for our understanding of
the nature of reality. Since some of the less mathematically oriented readers may have had some difficulties in
following our explanation of quantum mechanics by means of the quantum machine we shall now give a second, more
metaphorical and less technical, example of our creation-discovery view.

\section {Cracking walnuts and quantum mechanics.}

Consider the following experiment: `we take a walnut out of a basket, and break it open in order to eat it'. Let
us look closely at the way we crack the nut. We don't use a nutcracker, but simply take the nut between the palms
of our two hands, press as hard as we can, and see what happens. Everyone who has tried this knows that different
things can happen. A first possibility to envisage is that the nut is mildewed. If after cracking the shell the
walnut turns out to be mildewed, then we don't eat it.

\par Assume for a moment that the only property of the nut that plays a role in our eating it or not is the
property of being mildewed or not. Assume now that there are $N$ walnuts in the basket. Then, for a given nut $k$
(we have $1 \le k \le N$), there are always two possible results for our experiment: $E_1$, we crack the nut and
eat it (and then following our hypothesis, it was not mildewed); $E_2$, we crack the nut and don't eat it (and
then it was mildewed). Suppose that $M$ of the $N$ nuts in the basket are mildewed. Then the probability that our
experiment for a nut $k$ yields the result $E_1$ is given by the ratio ${{(N - M)}\over N}$, and that it yields
the result $E_2$, by $N\over M$. These probabilities are introduced by our lack of knowledge of the complete
physical reality for the nut. Indeed, the nut $k$ is either mildewed or not before we proceed to break it open.
Had we known about its being mildewed without having to crack the nut, then we could have eliminated the
probability statement, which is simply the expression of our lack of knowledge about the deeper unknown reality
of the nut. We could have selected the nuts for eating by removing from the basket all the mildewed ones The
classical probability calculus is based, as above, upon a priori assumptions as to the nature of existing
probabilities. 

\par Everyone who has had any experience in cracking walnuts knows that other things can happen. Sometimes, we
crush the nut upon cracking the shell. We then have to make an assessment of the damage incurred, and decide
whether or not it is worth while to try and separate out the nut from the fragments of the shell. If not, we
don't eat the nut. Taking into account this more realistic situation, we have to drop our hypothesis that the
only factor determining our eating the nut is the mildew, existing before the cracking. Now there are two
factors: the mildew, and the state of the nut `after' the act of cracking. Again we have two possible results for
our experiment: $E_1$, we don't eat the nut (then it was mildewed or is crushed upon cracking); and $E_2$, we eat
it (then it was not mildewed and cleanly cracked). For a given nut $k$ these two possible results will occur with
a certain probability. We perceive immediately that this sort of probability depends on the way we crack the nut,
and is thus of a different nature from the one only related to the presence of mildewed nuts. Before cracking the
nuts, there is no way of separating out those which will be cleanly cracked and those which will be crushed. This
distinction cannot be made because it is partly created by the cracking experiment itself. This is a nice example
of how aspects of physical reality can be created by the measurement itself, namely, the cracking open of the
walnuts, and it can be clearly understood why the probability that comes in by this effect is of another nature
and cannot be eliminated by looking for a deeper description of the entity under study.

\par We can state now easily our general creation-discovery view for the case of the nuts. The mildewed nature of
the nut is a property that the nut has before and independently of the fact that we break it. When we break the
nut and find out that it is mildewed, then this finding is a `discovery'. These discoveries, related to outcomes
of experiments, obey a classical probability calculus, expressing our lack of knowledge about something that was
already there before we made the experiment. The crushed or cleanly cracked nature of the nut is not a discovery
of the experiment of cracking. It is a creation. Indeed, depending on how we perform the experiment, and on all
other circumstantial factors during the experiment, some nuts will come out crushed, while others will be cleanly
cracked.

\par The mathematical structure of the probability model necessary to describe the probabilities for cleanly
cracked or crushed nuts is quite different from that needed for mildewed or non-mildewed ones. More specifically: 

\begin{quotation}
\noindent {\it The probability structure corresponding to the indeterminism resulting from a lack of knowledge of
an existing physical reality is a classical Kolmogorov probability model.} \end{quotation}
\begin{quotation}
\noindent {\it The
probability structure corresponding to the indeterminism resulting from the fact that during a measurement new
elements of physical reality, which thus did not exist before the measurement,  are created is a quantum-like
probability model.}
\end{quotation}

\noindent Quantum probabilities can thus be taken as resulting from a lack of knowledge of the interaction between
the measuring apparatus and the quantum entity during the measuring experiment. This interaction creates new
elements of physical reality which did not exist before the measurement. This is the explanation which we propose
to account for quantum probabilities.

We should point out that the non-Kolmogorovian nature of the probability model corresponding to situations of
creation cannot be shown for the case of a single experiment, as considered. At least three different experiments
with two outcomes of the creation type are necessary to prove in a formal way that a description within a
Kolmogorovian model is not possible. We refer to [7, 8, 9] for the details of such a proof for the quantum spin
1/2 model. The fact that we need at least three experiments does not however suppress the fact that the physical
origin of the non-Kolmogorovian behavior is clearly due to the presence of explicit creation aspects [52].

Let us now assume that we have removed all the mildewed walnuts from the basket. We then have the situation where
none of the nuts are mildewed. In the physicist's jargon we say that the individual nuts are in a pure state,
relative to the property of being mildewed or not. In the original situation when there were still mildewed nuts
present, an individual nut was in a mixed state, mildewed and not mildewed, with weighting factors $M\over N$ and
${(N - M)\over N}$. In the new situation with the basket containing only non-mildewed nuts, we consider an event
$m$: we take a non-mildewed walnut, and carry out the measurement consisting in cracking the nut. We have here
the two possible results: $E_1$, the nut is cleanly cracked  and we eat it; $E_2$, the nut is crushed and we
don't eat it. The result depends on what takes place during the cracking experiment. We therefore here introduce
the concept of potentiality. For the case of mildewed or non-mildewed nuts we could assert for each nut that,
previously to the experiment, the nut was mildewed or not. For the case of cleanly cracked or crushed nuts, we
cannot relate the outcome of the cracking to any anterior property of the walnut. What we can assert however is
that each walnut is potentially cleanly cracked (and will then be eaten), or potentially crushed (and then will
not be eaten).  

\par Nobody will have any difficulty in understanding the walnut example. What we propose is that one should try
to understand quantum reality in a similar manner. The only difference is that for the measurements in quantum
mechanics which introduce a probability of the second type (i.e. with the creation of a new element of physical
reality during the measurement), we find it difficult to visualize just what this creation is. This is the case
for instance for detection experiments of a quantum entity. Intuitively, we associate the detection process with
the determination of a spatial position which already exists. But now, we must learn to accept that the detection
of a quantum entity involves, at least partially, the creation of the position of the particle during the
detection process. Walnuts are potentially cleanly cracked or crushed, and likewise quantum entities are
potentially within a given region of space or potentially outside it. The experiment consisting in finding or not
finding a quantum entity in a given region takes place only after setting up in the laboratory the measuring
apparatus used for the detection, and it requires the interaction of the quantum entity with that measuring
apparatus. Consequently, the quantum entity is potentially present and potentially not present in the region of
space considered.

\par It will be observed that this description of quantum measurements makes it necessary to reconsider our
concept of space. If a quantum entity in a superposition state between two separated regions of space is only
potentially present in both of these region of space, then space is no longer the setting for the whole of
physical reality. Space, as we intuitively understand it, is in fact a structure within which classical relations
between macroscopic physical entities are established. These macroscopic entities are always present in space,
because space is essentially the structure in which we situate these entities. This need not be, and is not the
case for quantum entities. In its normal state, a quantum entity does not exist in space, it is only by means of
a detection experiment that it is, as it were, pulled into space. The action of being pulled into space
introduces a probability of the second type (the type associated with cracking the walnuts open), since the
position of the quantum entity is partially created during the detection process.

\par Let us consider now a neutron (photon) in Rauch's experiment (Wheeler's delayed-choice experiment) and let us
describe this situation within the creation-discovery view. We accept that the neutron (photon) while it travels
between the source and the detector is not inside space. It remains a single entity traveling through reality and
the two paths $n$ and $s$ are regions of space where the neutron (photon) can be detected more easily than in
other regions of space when a detection experiment is carried out. The detection experiment is considered to
contain explicitly a creation element and pulls the neutron (photon) inside space. If no detection experiment is
carried out, and no physical apparatuses related to this detection experiment are put into place, the neutron
(photon) is not traveling on one of the two paths $n$ or $s$. 

We can understand now how the `subjective' part of the Copenhagen interpretation disappears. In the creation-
discovery view the choice of the measurement, whether we choose to detect or to make an interference experiment,
does not influence the intrinsic nature of the quantum entity. In both choices the quantum entity is traveling
outside space, and the effect of an experiment appears only when the measurement related to the experiment
starts. If a detection measurement is chosen the quantum entity starts to get pulled into a place in space where
it localizes. If an interference experiment is chosen the quantum entity remains outside space, not localized,
and interacts from there with the macroscopic material apparatuses and the fields, and this interaction gives
rise to the interference pattern.

\section{Were do the quantum paradoxes go?}

We have analyzed in foregoing sections the manner in which the creation-discovery view resolves the problems that
are connected to the de Broglie theory and the Copenhagen interpretation. We would like to say now some words
about the quantum paradoxes. Our main conclusion  relative to the quantum paradoxes is the following: some are
due to intrinsic structural shortcomings of the orthodox theory, while others find their origin in the nature of
reality, and are due to the pre-scientific preconception about space that we have been able to explain. In this
way we can state that the generalized quantum theories together with the creation-discovery view resolve the
well-known quantum paradoxes. We do not have the space here to go into all the delicate aspects of the paradoxes,
and refer therefore to the literature. We shall however present a sufficiently detailed analysis of certain
cases, so that it becomes clear how the paradoxes are solved within the generalized quantum theories and the
creation-discovery view.

\smallskip
\noindent {\it (1) The measurement problem and Schr\"odinger's cat paradox :}
\smallskip

\noindent
If one tries to apply orthodox quantum mechanics to describe a system containing both a quantum entity and the
macroscopic measuring apparatus, one is led to very strange predictions. It was Schr\"odinger who discussed this
problem in detail, so let us consider the matter from the point of view of his cat [53]. Schr\"odinger imagined
the following thought experiment. He considered a room containing a radioactive source and a detector to detect
the radioactive particles emitted. In the room there is also a flask of poison and a living cat. The detector is
switched on for a length of time such there is exactly a probability 1/2 of detecting a radioactive particle
emitted by the source. Upon detecting a particle, the detector triggers a mechanism which breaks the flask,
liberating the poison and killing the cat. If no particle is detected, nothing happens, and the cat stays alive.
We can know the result of the experiment only when we go into the room to see what has happened. If we apply the
orthodox quantum formalism to describe the experiment (cat included), then, until the moment that we open the
door, the state of the cat, which we denote by $p_{cat}$, is a superposition of the two states ``the cat is dead",
written $p_{dead}$, and ``the cat is alive", written $p_{live}$. Thus, $p_{cat} = (p_{dead} + p_{live})/\sqrt{2}$. 

\par The superposition is suppressed, giving a change in the quantum mechanical state, only at the instant when we
go into the room to see what has taken place. We first want to remark that if we interpret the state as described
by the orthodox quantum mechanical wave function as a mathematical object giving exclusively {\it our knowledge}
of the system, then there would be no problem with Schr\"odinger's cat. Indeed, from the point of view of our
knowledge of the state, we can assume that before opening the door of the room the cat was already dead or was
still alive, and that the quantum mechanical change of state simply corresponds to the change in our knowledge of
the state. This {\it knowledge picture} would also resolve another problem. According to the orthodox quantum
formalism, the superposition state $p_{cat} = (p_{dead} + p_{live})/\sqrt{2}$ is instantaneously transformed, at
the instant when one opens the door, into one of the two component states $p_{dead}$ or $p_{live}$. This sudden
change of the state, which in the quantum mechanical jargon is called {\it the collapse of the wave function},
thus has a very natural explanation in the {\it knowledge picture}. Indeed, if the wave function describes our
knowledge of the situation, then the acquisition of new information, as for instance by opening a door, can give
rise to an arbitrarily sudden change of our knowledge and hence also of the wave function. 

\par The {\it knowledge picture} cannot however be correct, because it is a hidden variable theory. Indeed, the
quantum mechanical wave function does not describe the physical reality itself, which exists independently of our
knowledge of it, but describes only our knowledge of the physical reality. It would then follow, if the {\it
knowledge picture} is correct, that there must exist an underlying level of reality which is not described by a
quantum mechanical wave function. For the cat experiment, this underlying level describes the condition of the
cat, dead or alive, independently of the knowledge of this condition we acquire by entering the room. The {\it
knowledge picture} therefore leads directly to a {\it hidden variable theory}, where hidden variables describe
the underlying level of reality. As we mentioned already, it can be shown that a probabilistic theory, in which a
lack of knowledge of an underlying level of reality lies at the origin of the probabilistic description (a hidden
variable theory), always satisfies Kolmogorov's axioms. Now, the quantum mechanical theory does not satisfy these
axioms, so that the {\it knowledge picture} is necessarily erroneous. One also has direct experimental evidence,
in connection with the Bell inequalities, which confirms that any state-type hidden variable hypothesis is wrong. 

\par Hence, the quantum mechanical wave function represents not our knowledge of the system, but its real physical
state, independently of whether the latter is known or not. In that case, however, Schr\"odinger's cat presents
us with a problem. Is it really possible that, before the door of the room is opened, the cat could be in a
superposition state, neither living nor dead, and that this state, as a result of opening of the door, is
transformed into a dead or live state? It does seem quite impossible that the real world could react in this
manner to our observation of it. A physical reality such that its states can come into being simply because we
observe it, is so greatly in contradiction with all our real experience that we can hardly take this idea
seriously. Yet it does seem to be an unescapable consequence of orthodox quantum mechanics as applied to a global
physical situation, with macroscopic components.

In the new physical general description that we have proposed [29, 30, 31, 32] it is perfectly possible and even
very natural to make a distinction between different types of experiments. One will thus introduce the concept of
a {\it classical experiment}: this is an experiment such that, for each state $p$ of the entity $S$, there is a
well-determined result $x$. For a classical experiment, the result is fully predictable even before the
experiment is carried out. A collection $\cal E$ of relevant experiments will generally comprise both classical
and non-classical ones. It is possible to prove a theorem stating that the classical part of the description of
an entity can always be separated out [29, 31, 54]. The collection of all possible states for an entity can then
be expressed as the union of a collection of classical mixtures, such that each classical mixture is determined
by a set of non-classical micro-states. When we formulate within this general framework the axioms of quantum
mechanics, it can be shown that the set of states in a classical mixture can be represented by a Hilbert space.
The collection of all the states of the entity is then described by an infinite collection of Hilbert spaces, one
for each classical mixture. Orthodox quantum mechanics is in this formulation the limiting case for which no
classical measurement appears, corresponding effectively to the existence of a single Hilbert space. Classical
mechanics is the other limiting case, which is such that only classical measurements are present, and for which
the formulation corresponds to a phase space description. The general case for an arbitrary entity is neither
purely quantum nor purely classical, and can only be described by a collection of different Hilbert spaces. When
one considers the measuring process within this general formulation, there is no Schr\"odinger cat paradox.
Opening the door is a classical operation which does not change the state of the cat, and the state can thus also
be described within the general formulation, and the quantum collapse occurs when the radioactive particle is
detected by the detector, which is a non-classical process, also within the general description.

\par The general formalism provides more than the resolution of the Schr\"odinger cat paradox. It makes it
possible to consider quantum mechanics and classical mechanics as two particular cases of a more general theory.
This general theory is quantum-like, but introduces no paradoxes for the measuring process because one can treat,
within the same formalism, the measuring apparatus as a classical entity, and the entity to be measured as a
quantum entity. The paradoxes associated with measurements result from the structural limitations of the orthodox
quantum formalism. This decomposition theorem of a general description into an direct product of irreducible
descriptions, where each irreducible description corresponds to one Hilbert space, had been shown already within
the mathematical generalizations of quantum mechanics [20, 21]. The aim then was to give an explanation for the
existence of super-selection rules. The decomposition was later generalized for the physical formalisms [29, 31,
54].

\smallskip
\noindent {\it (2) The Einstein-Podolsky-Rosen paradox :}
\smallskip

\noindent The general existence of superposition states which lies at the root of the Schr\"odinger cat paradox,
was exploited by Einstein, Podolsky, and Rosen (EPR) to construct a far subtler paradoxical situation. EPR
consider the case of two separated entities $S_1$ and $S_2$, and the composite entity $S$ which these two entities
constitute. They show that it is always possible to bring the composite entity $S$ in a state in such a manner
that a measurement on one of the component entities determines the state of the other component entity. For
separated entities, this is a quantum mechanical prediction which contradicts the very concept of separateness.
Indeed, for separated entities the state of one of the entities can a priori not be affected by how one acts upon
the other entity, and this is confirmed by all experiments which one can carry out on separated entities.

\par Here again, we can resolve the paradox by considering the situation in the framework of the new general
formalism. There, one can show that a composite entity $S$, made up of two separated entities $S_1$ and $S_2$,
never satisfies the axioms of orthodox quantum mechanics, even if allowance is made for classical experiments as
was done in the case of the measurement paradox [11, 29, 30]. Two of the axioms of orthodox quantum mechanics
({\it weak modularity}, and the {\it covering law}) are never satisfied for the case of an entity $S$ made up of
two separated quantum entities $S_1$ and $S_2$. This failure of orthodox quantum mechanics is structurally much
more far-reaching than that relating to the measuring problem. There one could propose a solution in which the
unique Hilbert space of orthodox quantum mechanics is replaced by a collection of Hilbert spaces, and one remains
more or less within the framework of the Hilbert space formalism (this is the way that super-selection rules were
described even within one Hilbert space). The impossibility of describing separated entities in orthodox quantum
mechanics is rooted in the vector space structure of the Hilbert space itself. The two unsatisfied  axioms are
those associated with the vector space structure of the Hilbert space, and to dispense with these axioms, as is
required if we wish to describe separated entities, we must therefore construct a totally new mathematical
structure for the space of states [55, 56, 57]. 
	
\smallskip
\noindent {\it (3) Classical, quantum and intermediate structures.} 
\smallskip

\noindent To abandon the vector space structure for the collection $\Sigma$ of all possible states for an entity
is a radical mathematical operation, but recent developments have confirmed its necessity. The possibility of
accommodating within one general formalism both quantum and classical entities has resolved the measurement
paradox. If the quantum structure can be explained by the presence of a lack of knowledge on the measurement
process, as it is the case in our `hidden-measurement' approach, we can go a step further, and wonder what types
of structure arise when we consider the original models, with a lack of knowledge on  the measurement process,
and introduce a variation of the magnitude of this lack of knowledge.  We have studied the quantum machine under
varying `lack of knowledge', parameterizing  this variation by a number $\epsilon \in [0,1]$, such that $\epsilon
= 1$ corresponds to the  situation of maximal lack of knowledge, giving rise to a quantum structure, and
$\epsilon =  0$ corresponds to the situation of zero lack of knowledge, generating a classical  structure, and
other values of $\epsilon$ correspond to intermediate situations, giving rise to a  structure that is neither
quantum nor classical [4, 45, 46, 47, 58, 59, 60]. We have called this model the $\epsilon$-model, and we have
been able to proof that here again the same two axioms, weak modularity and the covering law, cannot be satisfied
for the intermediate situations - between quantum and classical [4, 45, 46, 47, 58, 59]. A new theory dispensing
with these two axioms would allow for the description not only of structures which are quantum, classical, mixed
quantum-classical, but also of intermediate structures, which are neither quantum nor classical. This is then a
theory for the mesoscopic region of reality, and we can now understand why such a theory could not be built
within the orthodox theories, quantum or classical.

\section{Standard quantum mechanics as a first order non classical theory.}

\noindent As our $\epsilon$ version of the quantum machine shows, there are different quantum-like theories
possible, all giving rise to quantum-like probabilities, that however differ numerically from the probabilities
of orthodox quantum mechanics. These intermediate theories may allow us to generate models for the mesoscopic
entities, and our group in Brussels is now investigating this possibility. The current state of affairs is the
following: quantum mechanics and classical mechanics are both extremal theories, corresponding relatively to a
situation with maximum lack of knowledge and a situation with zero lack of knowledge on the interaction between
measuring apparatus and the physical entity under study. Most real physical situations will however correspond to
a situation with a lack of knowledge of the interaction with the measuring apparatus that is neither maximal nor
zero, and as a consequence the theory describing this situation will have a structure that is neither quantum 
nor classical. It will be quantum-like, in the sense that the states are changed by the measurements, and that
there is a probability involved as in quantum mechanics, but the numerical value of this probability will be
different from the numerical value of the orthodox quantum mechanical probabilities. If this is the case, {\it
why does orthodox quantum mechanics has so much success, both in general and in its numerical predictions?} In
this section we want to suggest an answer to this question.  Let us consider the case of an entity $S$, and two
possible states $p_u$ and $p_v$ corresponding to this entity. We also consider all possible measurements that can
be performed on this entity $S$, with the only restriction that for each measurement considered it must be
possible that, when the entity is in the state $p_v$, it can be changed by the measurement into the state $p_u$.
Among these measurements there will be deterministic classical measurements, there will be quantum measurements,
but there will also be super-quantum measurements (giving rise to a probability greater than that predicted by 
quantum mechanics) and sub-quantum measurements (giving rise to a probability that lies between classical and
quantum predictions). All these different measurements are considered. We suppose now that we cannot distinguish
between these measurements, and hence the actual measurement that we perform, and which we denote $\Delta(u,v)$,
is a random choice between all these possible measurements. We shall call this measurement the `universal'
measurement connecting $p_v$ and $p_u$. We may remark that if we believe that there is `one' reality then also
there is only `one' universal measurement $\Delta(u,v)$ connecting $p_v$ and $p_u$. We now ask what is the
probability $P_\Delta(p_u, p_v)$ that by performing the universal measurement $\Delta(u,v)$, the state $p_v$ is
changed into the state $p_u$. 

\par There is a famous theorem in quantum mechanics that makes it possible for us to show that the universal
transition probability $P_\Delta(p_u, p_v)$ corresponding to a universal measurement $\Delta(u,v)$ connecting
states $p_u$ and $p_v$ is the quantum transition probability $P_q(p_u, p_v)$ connecting these two states $p_v$
and $p_u$. This is Gleason's theorem. 

Gleason's theorem proves that, for a given vector $u$ of a Hilbert space $\cal H$, of dimension at least $3$,
there exists only one probability measure $\mu_u$ on the set of closed subspaces of this Hilbert space, with
value $1$ on the ray generated by $u$, and this is exactly the probability measure used to calculate the quantum
transition probability from any state to this ray generated by $u$. Gleason's theorem is only valid for a Hilbert
space of dimension at least three. The essential part of the demonstration consists in proving the result for a
three-dimensional real Hilbert space. Indeed, the three-dimensional real Hilbert space case contains already all
the aspects that make Gleason's theorem such a powerful result. This is also the reason that we here restrict our
`interpretation' of Gleason's result to the case of a three dimensional real Hilbert space.
\begin{quotation}
\noindent {\it Theorem (Gleason)} : The only positive function $w(p_v)$ that is defined on the rays $p_v$ of a
three dimensional real Hilbert space $R^3$, and that has value $1$ for a given ray $p_u$, and
that is such that $w(p_x) + w(p_y) + w(p_z) = 1 $ if the three rays $p_x$, $p_y$, $p_z$ are mutually
orthogonal, is given by $w(p_v) = \vert < u , v > \vert^2 $
\end{quotation}   

\noindent Let us now consider two states $p_u$ and $p_v$, and a measurement $e$ (which is not a priori taken to be
a quantum measurement) that has three eigenstates $p_u$, $p_y$ and $p_z$, which means that it transforms any
state into one of these three states after the measurement. The probability $P_e(p_u, p_v)$, that the measurement
$e$ transforms the state $p_v$ into the state $p_u$ is given by a positive function $f(v, u, x, y)$ that can
depend on the four vectors $v, u, x$ and $y$. In the same way we have $P_e(p_x, p_v) = f(v, x, y, u)$,  $P_e(p_y,
p_v) = f(v, y, u, x)$, and $f(v, u, x, y) + f(v, x, y, u) + f(v, y, u, x) = 1$. This is true, independent of the
nature of the measurement $e$. If $e$ is a quantum measurement, then $f(v, u, x, y) = \vert < v , u > \vert^2$,
and the dependence on $x$ and $y$ disappears, because the quantum transition probability only depends on the
state before the measurement and the eigen state of the measurement that is actualized, but not on the other
eigenstates of the measurement. Gleason's theorem states that `if the transition probability depends only on the
state before the measurement and on the eigenstate of the measurement that is actualized after the measurement,
then this transition probability is equal to the quantum transition probability'. But this Gleason property
(dependence of the transition probability only on the state before the measurement and the eigenstate that is
actualized after the measurement) is precisely a property that is satisfied by what we have called the
`universal' measurements. Indeed, by definition, the transition probability for a universal measurement only
depends on the state before the measurement and the actualized state after the measurement. Hence Gleason's
theorem shows that the transition probabilities connected with universal measurements are quantum mechanical
transition probabilities.

We now go a step further and proceed to interpret the quantum measurements as if they are universal measurements.
This means that quantum mechanics is taken to be the theory that describes the probabilistics of possible
outcomes for measurements which are mixtures of all imaginable types of measurements. Quantum mechanics is then
the first order non-classical theory. It describes the statistics that goes along with a random choice between
any arbitrary type of manipulation that changes the state $p_v$ of the system under study into the state $p_u$,
in such a way that we know nothing of the mechanism of this change of state. The only information we have is that
`possibly' the state before the measurement, namely $p_v$, is changed into a state after the measurement, namely
$p_u$. If this is a correct explanation for quantum statistics, it accounts for its success in so many regions of
reality, both in general and also for its numerical predictions.

\section{Relativity theory: is reality vanishing?}

When James Clerck Maxwell developed his field theory for electromagnetic radiation the seeds were sown of a
problem of the {\it `the classical mechanical view'}. Indeed, while the classical mechanical equations
are invariant for Galilean transformations - this invariance expresses mathematically an additional intuition
within our intuitive view on reality, namely, that the laws of physics remain the same in another coordinate
system moving relatively to us with constant velocity -  Maxwell's equations turn out to be invariant for a
totally different type of transformations. The problem was recognized by Hendrik Antoon Lorentz - hence the name
'Lorentz transformation' given to this new set of transformations - as also by Henri Poincar\'e and others,
around the turn of the century. As the story goes, the young Albert Einstein also pondered on this problem as a
physics student, and his reflection was at the origin of the article in which he formulated the theory of
relativity [61].

In relativity theory a very subtle but straightforward fundamental subjective element is introduced within the
nature of reality itself. It is well recognized in broad circles that the meaning of quantum mechanics as related
to the nature of reality has not yet been understood. For relativity theory there seems to be however a common
belief, and this is certainly partly due to the straightforward operational manner in which the theory was
introduced by Albert Einstein, that its consequences for the nature of reality have been well understood by the
specialists. As our analysis will show - and contrary to what is believed by many physicists - the profound
meaning of relativity theory for the nature of reality has not yet been understood at all.

Usually relativity theory is introduced with a seemingly very well defined ontological basis [62]. The collection
of events, each event parametrized by four real numbers $(x_0, x_1, x_2, x_3)$, is considered to be the basic
structure of the theory. For a particular observator connected to a particular reference frame, there is no
problem of how to use this four-dimensional time-space manifold scheme to decide what `his personal reality' is.
His personal reality is indeed the `space-cut' that his reference frame makes with the four-dimensional
time-space manifold. This space-cut, however, only determines a reality connected to a particular reference
frame, and at first sight it is not possible to put together the space-cuts of different reference frames in such
a manner that they form one reality. All this is well known, and this problem was in fact already at the origin
of the construction of special relativity in the original paper by Albert Einstein, namely his critique on the
concept of simultaneity [61]. 

But there is a fundamental problem in relativity theory in relation to the question: ``What is reality?". Sometimes
the statement is made rather vaguely and never with a sound conceptual basis, that reality in relativity theory
`is' the four-dimensional time-space continuum. But if this position is taken, there is another major conceptual
problem: indeed then there is no change and no evolution in time. Eventually we could still accept that material
reality would be frozen in four dimensions, but then the question remains: what are we? I myself, and I suppose
also all of you readers, am convinced of the fact that I am not my past and my future. I am now. In this way,
relativity theory conflicts with our deep intuition about the nature of reality in
a manner such that we can not even well identify just where the contradiction lies. We have analysed in great
detail this situation in [63, 64], and shall come back to it after introducing an operational definition for
reality such that we can detect what is the `real' mystery.

\smallskip
\noindent {\it (1) Experiences.}
\smallskip

\noindent The basic concept in our analysis of the operational foundation of reality is that of an experience.
An experience is the interaction between a participator and a piece of the world. When the participator lives
such an experience, we shall say that this experience is {\it present}, and we shall call it the {\it present
experience} of the participator. We remark that we consciously use the word `participator' instead of the word
`observer' to indicate that we consider the cognitive receiver to participate creatively in his cognitive act.
When we consider a measurement, then we conceive that for this situation the experimentator and his experimental
apparatus together constitute the participator, and that the physical entity under study is the piece of the
world that interacts with the participator. The experiment is the experience.

Let us give some examples of experiences. Consider the following situation: I am inside my house in Brussels. It
is night, the windows are shut. I sit in a chair, reading a novel. I have a basket filled with walnuts at my
side, and from time to time I take one of them, crack it and eat it. My son is in bed and already asleep. New
York exists and is busy.

Let us enumerate the experiences that are considered in such a situation:

\noindent (1) $E_1$(I read a novel)

\noindent (2) $E_2$(I experience the inside of my house in
Brussels)

\noindent (3) $E_3$(I experience that it is night)

\noindent (4) $E_4$(I take a walnut, crack it and eat it)

\noindent (5) $E_5$(I see that my son is in bed and asleep)

\noindent (6) $E_6$(I experience that New York is busy)

The first very important remark I want to make is that obviously I do not experience all these experiences at
once. On the contrary, in principle, I only experience one experience at once, namely my present experience. Let
us suppose that my present experience is $E_1$(I read a novel). Then a lot of other things happen while I am
living this present experience. These things happen in my present reality. While 'I am reading the novel' some of
the happenings that happen are the following: $H_1$(the novel exists), $H_2$(the inside of my house in Brussels
exists), $H_3$(it is night), $H_4$(the basket and the walnuts exist, and are at my side), $H_5$(my son is in bed
and is sleeping), $H_6$(New York exists and is busy). All the happenings, and much more, happen while I live the
present experience $E_1$(I read a novel).

Why is the structure of reality such that what I am just saying is evident for everybody (and therefore
shows that we are not conscious of the structure and construction that is behind this evidence)? 

Certainly it is not because I experience also these other happenings. My only {\it present} experience is the
experience of reading the novel. But, and this is the origin of the specific structure and construction of
reality, I could have chosen to live an experience including one of the other happenings {\it in replacement} of
my present experience. Let me recapitulate the list of the experiences that I could have chosen to experience in
replacement of my present experience: $E_2$(I observe that I am inside my house in Brussels), $E_3$(I see that it
is night),
$E_4$(I take a walnut, crack it and eat it), $E_5$(I go and look in the bedroom to see that my son is asleep),
$E_6$(I take the plane to New York and see that it is busy).  This example indicates how reality is structured
by us. 

First of all we have tried to identify two main aspects of an experience. The aspect that is controlled and
created by me, and the aspect that just happens to me and can only be known by me. Let us introduce this
important distinction in a formal way. 

\smallskip
\noindent {\it (2) Creations and happenings.}
\smallskip

\noindent
To see what I mean, let us consider the experience $E_4$(I take a walnut, crack it and eat it). In this
experience, there is an aspect that is an action  of me, the taking and the cracking, and the eating. There is
also an aspect that is an observation of me, the walnut and the basket. By studying how our senses work, I can
indeed say that it is the light reflected on the walnut, and on the basket, that gives me the experience of
walnut and the experience of basket. This is an explanation that only now can be given; it is, however, not what
was known in earlier days when the first world-models of humanity were constructed. But without knowing the
explanation delivered now by a detailed analysis, we could see very easily that an experience contains always two
aspects, a {\it creation}-aspect, and an {\it observation}-aspect, simply because our will can only control part
of the experience. This is the creation-aspect.

For example, in $E_1$(I read a novel) the reading is created by me, but the novel is not created by me. In general
we can indicate for an experience the aspect that is created by me and the aspect that is not created by me. The
aspect not created by me lends itself to my creation. We can reformulate an experience in the following way:
$E_4$(I take a walnut, crack it and eat it) becomes $E_4$(The walnut is taken by me, and lends itself to my
cracking and eating) and $E_1$(I read a novel) becomes $E_1$(The novel lends itself to my reading).  The taking,
cracking, eating, and reading will be called {\it creations} or actions and will be denoted by $C_4$(I take,
crack and eat) and $C_1$(I read). The walnut and the novel will be called {\it happenings} and will be denoted by
$H_4$(The walnut) and $H_7$(The novel).

\begin{quotation} \noindent {\it A creation is that aspect of an experience
created, controlled, and acted upon by me, and a happening is that
aspect of an experience lending itself to my creation, control
and action.}
\end{quotation}
  
\noindent An experience is determined by a description of the creation and a description of the
happening. Creations are often expressed by verbs: to take, to crack, to eat, and to read, are the verbs that
describe my creations in the examples. The walnut and the novel are happenings that have the additional property
of being objects, which means happening with a great stability. Often happenings are expressed by a substantive.

\begin{quotation} \noindent {\it Every one of my experiences $E$ consists of one of
my creations $C$ and one of my happenings $H$, so we can write $E =
(C,H)$.}
\end{quotation}

\noindent A beautiful image that can be used as a metaphor for our model of the world is the image of the skier. A skier skis downhill. At every instant he or she has to be in complete harmony with the form of the mountain under-neath. The mountain is the happening. The actions of the skier are the creation. The skier's creation, in harmony fused with the skier's happening, is his or her experience.

\smallskip
\noindent { (3) \it The structure and construction of reality, present, past and
future.}
\smallskip

\noindent 
Let us again consider the collection of experiences: $E_1$(I read a novel), $E_2$(I observe that I am inside my
house in Brussels), E$_3$(I see that it is night), E$_4$(I take a walnut, crack it and eat it), E$_5$(I go and
look in the bedroom to see that my son is asleep) and E$_6$(I take the plane to New York and see that it is
busy). Let us now represent the  structure and construction of reality that is made out of this small collection
of experiences.

$E_1$(I read a novel) is my present experience. In my past I could, however, at several moments have chosen to do
something else and this choice would have led me to have another present experience than $E_1$(I read a novel).
For example:  One minute ago I could have decided to stop reading and observe that I am inside the house. Then
$E_2$(I observe that I am inside my house in Brussels) would have been my present experience.  Two minutes ago I
could have decided to stop reading and open the windows and see that it is night. Then $E_3$(I see that it is
night) would have been my present experience.  Three minutes ago I could have decided to stop reading, take a
walnut from the basket, crack it, and eat it. Then $E_4$(I take a walnut, crack it and eat it) would have been my
present experience.  Ten minutes ago I could have decided to go and see in the bedroom whether my son is asleep.
Then $E_5$(I go and look in the bedroom to see that my son is asleep) would have been my present experience.

Ten hours ago I could have decided to take a plane and fly to New York and see how busy it was. then $E_6$(I go to
New York and see that it is busy) would have been my present experience. 

\begin{quotation}
\noindent
{\it Even when they are not the happening aspect of my present experience, happenings 'happen' at present if they
are the happening aspect of an experience that I could have lived in replacement of my present experience, if I
had so decided in my past.}
\end{quotation}

\noindent The fact that a certain experience $E$ consisting of a creation $C$ and an happening $H$ is for me a
possible present experience depends on two factors:

\medskip
\noindent
(1) I have to be able to perform the creation.

\smallskip
\noindent
(2) The happening has to be available.

\medskip
\noindent For example, the experience $E_2$(I observe that I am inside my house in Brussels) is a possible
experience for me, if:

\medskip
\noindent
(1) I can perform the creation that consists in observing the inside of my house in Brussels. In other words, if
this creation is in my personal power.

\smallskip
\noindent
(2) The happening 'the inside of my house in Brussels' has to be available to me. In other words, this happening
has to be contained in my personal reality.

\begin{quotation}
\noindent
{\it The collection of all creations that I can perform at the present I will call my present personal power. The
collection of all happenings that are available to me at the present I will call my present personal reality.}
\end{quotation}

\noindent
I define as my present personal reality the collection of these happenings, the collection of happenings that are
available to one of my creations if I had used my personal power in such a way that at the present I fuse one of
these creations with one of these happenings.

\begin{quotation}
\noindent
{\it My present personal reality consists of all happenings that are available to me at present. My past reality
consists of all happenings that were available to me in the past. My future reality consists of all happenings
that will be available to me in the future. My present personal power consists of all creations that I can
perform at present. My past personal power consists of all the creations that I could perform in the past. My
future personal power consists of all creations I shall be able to perform in the future.}
\end{quotation}

\noindent
Happenings can happen 'together and at once', because to happen a happening does not have to be part of my present
experience. It is sufficient that it is available, and things can be available simultaneously. Therefore,
although my present experience is only one, my present personal reality consists of an enormous amount of
happenings all happening simultaneously.

This concept of reality is not clearly understood in present physical theories. Physical theories know how to
treat past, present and future. But reality is a construction about the possible. It is a construction about the
experiences I could have lived but probably will never live.

\smallskip
\noindent{\it (4) Material time and material happenings.}
\smallskip

\noindent 
From ancient times humanity has been fascinated by happenings going on in the sky, the motion of the sun, the
changes of the moon, the motions of the planets and the stars. These happenings in the sky are {\it periodic}. By
means of these periodic happenings humans started to {\it coordinate} the other experiences. They introduced the
counting of the years, the months and the days. Later on watches were invented to be able to coordinate
experiences of the same day. And in this sense {\it material} time was introduced in the reality of the human
species. Again we want to analyze the way in which this material time was introduced, to be able to use it
operationally if later on we analyze the paradoxes of time and space.

My present experience is seldom a material time experience. But in replacement of my present experience, I always
could have consulted my watch, and in this way live a material time experience $E_7$(I consult my watch and read
the time). In this way, although my present experience is seldom a material time experience, my present reality
always contains a material time happening, namely the happening $H_7$(The time indicated by my watch), which is
the happening to which the creation $C_7$(I consult) is fused to form the experience $E_7$.

We can try to use our theory for a more concrete description of that layer of reality that we shall refer to as
the layer of 'material or energetic happenings'. We must be aware of the fact that this layer is a huge one, and
so first of all we shall concentrate on those happenings that are related to the interactions between what we
call material (more generally energetic) entities. We have to analyze first of all in which way the
four-dimensional manifold that generally is referred to as the 'time-space' of relativity theory, is related to
this layer of material or energetic reality. We shall take into account in this analysis the knowledge that we
have gathered about the reality of quantum entities in relation with measurements of momentum and position.

\section {The structure and construction of reality and relativity theory.}

We consider the set of all material or energetic happenings and denote this set by ${\cal M}$. Happenings of
${\cal M}$ we shall denote by $m, n, o$. Let us consider such a happening $m$ that corresponds to a quantum
entity. Then this happening is characterized by the fact that it is always accessible to a creation of
localization (consisting in localizing the particle in a certain region of space), let us denote such a creation
of localization by $l$. Then the experience $(l, m)$ is an experience that can be parametrized by the coordinates
of a certain point $(x_0, x_1, x_2, x_3)$ of the four dimensional manifold that is referred to as time-space. 

However instead of performing a creation of localization, one can choose to perform a creation that consists in
measuring the momentum of the quantum entity. Let us denote this creation by $i$, then the happening $(i, m)$ can
be parametrized by the coordinates of a certain point $(p_0, p_1, p_2, p_3)$, that can be interpreted as the
four-momentum of general relativity theory.

We know from quantum theory that the quantum entity can be in different states, all corresponding to a different
statistics as related to repeated localizations and measurements of momentum. Let us denote these states by $q,
p, ...$. The quantum entity can be in an eigenstate $q(x_0, x_1, x_2, x_3)$ of position, which means that the
creation of localization in this eigenstate leads with certainty to a finding of the quantum entity in the point
$(x_0, x_1, x_2, x_3)$. The quantum entity can also be in an eigenstate $p(p_0, p_1, p_2, p_3)$ of momentum,
which means that by a measurement of momentum the entity will be found to have the momentum $(p_0, p_1, p_2,
p_3)$. But in general the quantum entity will be in a state that is neither an eigenstate of position nor an
eigenstate of momentum. It is only after the happening $p$ (the state of the quantum entity) has been fused with
one of the creations $l$ (the localization measurement) or $i$ (the momentum measurement) that will be in an
eigenstate of localization (a point of time-space) or of momentum (a point of four-momentum space). This is the
general situation for material happenings.

To show what are the problems that we can solve by means of our framework, we will concentrate now on the question
'what is reality in relativity theory?'. Since we have an operational definition of reality in our framework, we
can investigate this problem in a rigorous way.

Let us suppose that I am here and now in my house in Brussels, and it is June 1, 1996, 3 pm exactly. I want to
find out 'what is the material reality for me now?'. Let us use the definition of reality given in the foregoing
section and consider a place in New York, for example at the entrance of the Empire State building, and let us
denote, the center of this place by $(x_1, x_2, x_3)$. I also choose now a certain time, for example June 1,
1996, 3 pm exactly, and let me denote this time by $x_0$. I denote the happening that corresponds with the spot
$(x_1, x_2, x_3)$ located at the entrance of the Empire State building, at time $x_0$ by $m$. I can now try to
investigate whether this happening $m$ is part of my personal material reality. The question I have to answer is,
can I find a creation of localization $l$, in this case this creation is just the observation of the spot $(x_1,
x_2, x_3)$ at the entrance of the Empire State building, at time $x_0$, that can be fused with this happening
$m$. The answer to this question can only be investigated if we take into account the fact that I, who want to
try to fuse a creation of localization to this happening, am bound to my body, which is also a material entity. I
must specify the question introducing the material time coordinate that I coordinate by my watch. So suppose that
I coordinate my body by the four numbers $(y_0, y_1, y_2, y_3)$, where $y_0$ is my material time, and $(y_1, y_2,
y_3)$ is the center of mass of my body. We apply now our operational definition of reality. A this moment, June
1, 1996 at 3 pm exactly, my body is in my house in Brussels, which means that $(y_0, y_1, y_2, y_3)$ is a point
such that $y_0$ equals June 1, 1996, 3 pm, and $(x_1, x_2, x_3)$ is a point, the center of mass of my body,
somewhere in my house in Brussels. This shows that $(x_0, x_1, x_2, x_3)$ is different from $(y_0, y_1, y_2,
y_3)$, in the sense that $(x_1, x_2, x_3)$ is different form $(y_1, y_2, y_3)$ while $x_0 = y_0$.

The question is now whether $(x_0, x_1,  x_2, x_3)$ is a point of my material reality, hence whether it makes
sense to me to claim that now, June 1, 1996, 3 pm, the entrance of the Empire State building 'exists'. If our
theoretical framework corresponds in some way to our pre-scientific construction of reality, the answer to the
foregoing question should be affirmative. Indeed, we all believe that 'now' the entrance of the Empire State
building exists. Let us try to investigate in a rigorous way this question in our framework. We have to verify
whether it was possible for me to decide somewhere in my past, hence before June 1, 1996, 3 pm, to change some of
my plans of action, such that I would decide to travel to New York, and arrive exactly at June 1, 1996, 3 pm at
the entrance of the Empire State building, and observe the spot $(x_1, x_2, x_3)$. There are many ways to realize
this experiment, and we will not here go into details, because we shall come back later to the tricky parts of
the realization of this experiment. I could thus have experienced the spot $(x_1, x_2, x_3)$ at June 1, 1996, 3
pm, if I had decided to travel to New York at some time in my past. Hence $(x_0, x_1, x_2, x_3)$ is part of my
reality. It is sound to claim that the entrance of the Empire State building exists right now. And we note that
this does not mean that I have to be able to experience this spot at the entrance of the Empire State building
now, June 1, 1996, 3 pm, while I am inside my house of Brussels. I repeat again, reality is a construction about
the possible happenings that I could have fused with my actual creation. And since I could have decided so in my
past, I could have been at the entrance of the Empire State building, now, June 1, 1996, 3 pm.

Until this moment one could think that our framework only confirms our intuitive notion of reality, but our next
example shows that this is certainly not the case. Let us consider the same problem as above, but for another
point of time-space. We consider the point $(z_0, z_1, z_2, z_3)$, where $(z_1, z_2, z_3) = (x_1, x_2, x_3)$,
hence the spot we envisage is again the entrance of the Empire State building, and $z_0$ is June 2, 1996, 3 pm
exactly, hence the time that we consider is, tomorrow 3 pm. If I ask now first, before checking rigorously by
means of our operational definition of reality, whether this point $(z_0, z_1, z_2, z_3)$ is part of my present
material reality, the intuitive answer here would be 'no'. Indeed, tomorrow at the same time, 3 pm, is in the
future and not in the present, and hence it is not real, and hence no part of my present material reality (this
is the intuitive reasoning). If we go now to the formal reasoning in our framework, then we can see that the
answer to this question depends on the interpretation of relativity theory that we put forward. Indeed, let us
first analyze the question in a Newtonian conception of the world to make things clear. Remark that in a
Newtonian conception of the world (which has been proved experimentally wrong, so here we are just considering it
for the sake of clarity), my present material reality just falls together with 'the present', namely all the
points of space that have the same time coordinate June 1, 1996, 3 pm. This means that the entrance of the Empire
State building tomorrow 'is not part of my present material reality'. The answer is here clear and in this
Newtonian conception, my present personal reality is just the collection of all $(u_0, u_1, u_2, u_3)$ where $u_0
= y_0$ and $(u_1, u_2, u_3)$ are arbitrary. The world is not Newtonian, this we now know experimentally; but if
we put forward an ether theory interpretation of relativity theory (let us refer to such an interpretation as a
Lorentz interpretation) the answer again remains the same. In a Lorentz interpretation, my present personal
reality coincides with the present reality of the ether, namely all arbitrary points of the ether that are at
time $y_0$, June 1, 1996 3 pm, and again tomorrow the entrance of the Empire State building is not part of my
present material reality.

For an Einsteinian interpretation of relativity theory the answer is different. To investigate this I have to ask
again the question of whether it would have been possible for me to have made a decision in my past such that I
would have been able to make coincide $(y_0, y_1, y_2, y_3)$ with $(z_0, z_1, z_2, z_3)$. The answer here is that
this is very easy to do, because of the well known, and experimentally verified, effect of 'time dilatation'.
Indeed, it would for example be sufficient that I go back some weeks in my past, let us say April 1, 1996, 3 pm,
and then decide to step inside a space ship that can move with almost the speed of light, so that the time when I
am inside this space ship slows down in such a way, that when I return with the space ship to planet earth, still
flying with a speed close to the velocity of light, I arrive in New York at the entrance of the Empire State
building with my personal material watch indicating June 1, 1996 3 pm, while the watch that remained at the
entrance of the Empire State building indicates June 2, 1996 3 pm. Hence in this way I make coincide $(y_0, y_1,
y_2, y_3)$ with $(z_0, z_1, z_2, z_3)$, which proves that $(z_0, z_1, z_2, z_3)$ is part of my present material
reality. First I could remark that in practice it is not yet possible to make such a flight with a space ship.
But this point is not crucial for our reasoning. It is sufficient that we can do it in principle. We have not yet
made this explicit remark, but obviously if we have introduced in our framework an operational definition for
reality, then we do not have to interpret such an operational definition in the sense that only operations are
allowed that actually, taking into account the present technical possibilities of humanity, can be performed. If
we were to advocate such a narrow interpretation, then even in a Newtonian conception of the world, the star
Sirius would not exist, because we cannot yet travel to it. What we mean with operational is much wider. It must
be possible, taking into account the actual physical knowledge of the world, to conceive of a creation that can
be fused with the happening in question, and then this happening pertains to our personal reality.

\smallskip
\noindent {\it (1) Einstein versus Lorentz: has reality four
dimensions?}
\smallskip

\noindent
We can come now to one of the points that we want to make in this paper, clarifying the time paradox that
distinguishes an ether interpretation of relativity (Lorentz) from an Einsteinian interpretation. To see clearly
in this question, we must return to the essential aspect of the construction of reality in our framework, namely,
the difference between a creation and a happening. We have to give first another example to be able to make clear
what we mean.

Suppose that I am a painter and I consider again my present material reality, at June 1, 1996, 3 pm, as indicated
on my personal material watch. I am in my house in Brussels and let us further specify: the room where I am is my
workshop, surrounded by paintings, of which some are finished, and others I am still working on. Clearly all
these paintings exist in my presents reality, June 1, 1996, 3 pm. Some weeks ago, when I was still working on a
painting that now is finished, I could certainly have decided to start to work on another painting, a completely
different one, that now does not exist. Even if I could have decided this some weeks ago, everyone will agree
that this other painting, that I never started to work on, does not exist now, June 1, 1996, 3 pm. The reason for
this conclusion is that the making of a painting is a 'creation' and not a happening. It is not so that there is
some 'hidden' space of possible paintings such that my choice of some weeks ago to realize this other painting
would have made me to detect it. If this were to be the situation with paintings, then indeed also this painting
would exist now, in this hidden space. But with paintings this is not the case. Paintings that are not realized
by the painter are potential paintings, but they do not exist.

With this example of the paintings we can explain very well the difference between Lorentz and Einstein. For an
ether interpretation of relativity the fact that my watch is slowing down while I decide to fly with the space
ship nearly at the speed of light and return to the entrance of the Empire State building when my watch is
indicating June 1, 1996, 3 pm while the watch that remained at the Empire State building indicates June 2, 1996,
3 pm, is interpreted as a 'creation'. It is seen as if there is a real physical effect of creation on the
material functioning of my watch while I travel with the space ship, and this effect of creation is generated by
the movement of the space ship through the ether. Hence the fact that I can observe the entrance of the Empire
State building tomorrow June 2, 1996 3 pm, if had decided some weeks ago to start traveling with the space ship,
only proves that the entrance of the Empire State building tomorrow is a potentiality. Just like the fact that
this painting that I never started to paint could have been here in my workshop in Brussels is a potentiality.
This means that as a consequence the spot at the entrance of the Empire State building tomorrow is not part of my
present reality, just as the possible painting that I did not start to paint is not part of my present reality.
If we however put forward an Einsteinian interpretation of relativity, then the effect on my watch during the
space ship travel is interpreted in a completely different way. There is no physical effect on the material
functioning of the watch - remember that  most of the time dilatation takes place not during the accelerations
that the space ship undergoes during the trip, but during the long periods of flight with constant velocity
nearly at the speed of light - but the flight at a velocity close to the speed of light 'moves' my space ship in
the time-space continuum in such a way that time coordinates and space coordinates get mixed. This means that the
effect of the space-ship travel is an effect of a voyage through the time-space continuum, which brings me at my
personal time of June 1, 1996, 3 pm at the entrance of the Empire State building, where the time is June 2, 1996,
3 pm. And hence the entrance of the Empire State building is a happening, an actuality and not just a
potentiality, and it can be fused with my present creation.  This means that the happening $(z_0, z_1, z_2, z_3)$
of June 2, 1996, 3 pm, entrance of the Empire State building, is an happening that can be fused with my creation
of observation of the spot around me at June 1, 1996, 3 pm. Hence it is part of my present material reality. The
entrance of the Empire State building at June 2, 1996, 3 pm exists for me today, June 1, 1996 3 pm.

If we advocate an Einsteinian interpretation of relativity theory we have to conclude from the foregoing section
that my personal reality is four dimensional. This conclusion will perhaps not amaze those who always have
considered the time-space continuum of relativity as representing the new reality. Now that we have however
defined very clearly what this means, we can start investigating the seemingly paradoxical conclusions that are
often brought forward in relation with this insight.

\smallskip
\noindent { \it  (2) The process view confronted with the geometric
view.}
\smallskip

\noindent
The paradoxical situation that we can now try to resolve is the confrontation of the process view of reality with
the geometric view. It is often claimed that an interpretation where reality is considered to be related to the
four-dimensional time-space continuum contradicts another view of reality, namely the one where it is considered
to be of a process-like nature. By means of our framework we can now understand exactly what these two views
imply and see that there is no contradiction. Let us repeat now what in our framework is the meaning of the
conclusion that my personal reality is four dimensional. It means that, at a certain specific moment, that I call
my 'present', the collection of places that exist, and that I could have observed if I had decided to do so in my
past, has a four-dimensional structure, well represented mathematically by the four dimensional time-space
continuum. This is indeed my present material reality. This does not imply however that this reality is not
constantly changing. Indeed it is constantly changing. New entities are created in it and other entities
disappear, while others are very stable and remain into existence. This in fact is the case in all of the four
dimensions of this reality. Again I have to give an example to explain what I mean. We came to the conclusion
that now, at June 1, 1996, 3 pm the entrance of the Empire State building exists for me while I am in my house in
Brussels. But this is not a statement of derterministic certainty. Indeed, it is quite possible that by some
extraordinary chain of events, and without me knowing of these events, that the Empire State building had been
destructed; thus my statement about the existence of the entrance of the Empire State building 'now', although
almost certainly true, is not deterministically certain. The reason is again the same, namely that reality is a
construction of what I would have been able to experience, if I had decided differently in my past. The knowledge
that I have about this reality is complex and depends on the changes that go on continuously in it. What I know
from experience is that there do exist material objects, and the Empire State building is one of them, that are
rather stable, which means that they remain in existence without changing too much. To these stable objects,
material objects but also energetic fields, I can attach the places from where I can observe them. The set of
these places has the structure of a four-dimensional continuum. At the same time all these objects are
continuously changing and moving in this four-dimensional scenery. Most of the objects that I have used to shape
my intuitive model of reality are the material objects that surround us here on the surface of the earth. They
are all firmly fixed in the fourth dimension (the dimension indicated by the 0 index, and we should not call it
the time dimension) while they move easily in the other three dimensions (those indicated by the 1, 2, and 3
index). Other objects, for example the electromagnetic fields, have a completely different manner of being and
changing in this four-dimensional scenery. This means that in our framework there is no contradiction between the
four- dimensionality of the set of places and the process-like nature of the world. When we come to the
conclusion that the entrance of the Empire State building, tomorrow, June 2, 1996, 3 pm also exists for me now,
then our intuition reacts more strongly to this statement, because intuitively we think that this implies that
the future exists, and hence is determined and hence no change is possible. This is a wrong conclusion which
comes from the fact that during a long period of time we have had the intuitive image of a Newtonian present, as
being completely determined. We have to be aware of the fact that it is the present, even in the Newtonian sense,
which is not determined at all. We can only say that the more stable entities in our present reality are more
strongly  determined to be there, while the places where they can be are always there, because these places are
stable with certainty.

\smallskip
\noindent{ \it (3) The singularity of the reality construction.}
\smallskip

\noindent
We now come back to the construction of reality in our framework which we have confronted here with the
Einsteinian interpretation of relativity theory. Instead of wondering about the existence of the entrance of the
Empire State building tomorrow, June 2, 1996, 3 pm, I can also question the existence of my own house at the same
place of the time-space continuum. Clearly I can make an analogous reasoning and come then to the conclusion that
my own house, and the chair where I am sitting while reading the novel, and the novel itself, and the basket of
wall nuts beside me, etc..., all exist in my present reality at June 2, 1996 3 pm, hence tomorrow. If we put it
like that, we are even more sharply confronted with a counter-intuitive aspect of the Einsteinian interpretation
of relativity theory. But in our framework, it is a correct statement . We have to add however that all these
objects that are very close to me now June 1, 1996, 3 pm, indeed also exist in my present reality at June 2,
1996, 3 pm, but the place in reality where I can observe them is of course much further away for me. Indeed, to
be able to get there, I have to fly away with a space ship at nearly the velocity of light. We now come to a very
peculiar question that will confront us with the singularity of our reality construction. Where do I myself
exist? Do I also exist tomorrow June 2, 1996, 3 pm? If the answer to this question is affirmative, we would be
confronted with a very paradoxical situation. Because indeed I, and this counts for all of you also, cannot
imagine myself to exist at different instants of time. But our framework clarifies this question very easily. It
is impossible for me to make some action in my past such that I would be able to observe myself tomorrow June 2,
1996 3 pm. But if I had chosen to fly away and come back with the space-ship, it would be quite possible for me
to observe now, on June 1, 1996, at 3 pm on my personal watch, the inside of my house tomorrow June 2, 1996, 3
pm. As we remarked previously, this proves that the inside of my house tomorrow is part of my personal reality
today. But I will not find myself in it. Because to be able to observe my house tomorrow June 2, 1996 3 pm, I
have had to leave it. Hence, in this situation I will enter my house, being myself still at June 1, 1996, 3 pm,
but with my house and all the things in it, being at June 2, 1996, 3 pm. This shows that there is no
contradiction.

\section{What about the nature of reality?}

Let us finally investigate what is the meaning of all this for the nature of reality. As we remarked already in
our formal analysis of the construction of our personal reality, our most primitive intuition about the nature of
reality is that of a situation where there is {\it 'the past'}, {\it 'the present'} and {\it 'the future'}. {\it
'Reality'} is what {\it 'exists'} in {\it 'the present'} and is constantly changing, and new things are coming
into existence. {\it 'The past'} is a collection of what has been real, but does not exist anymore, while {\it
'the future'} is the field where the potentialities for possible realities are imagined by us. Let us refer to
this intuitive hypothesis about the nature of reality as the {\it 'past-present-future hypothesis'}.  Further we
think that reality consists of {\it 'entities'} and {\it 'interactions} between these entities, which exist at
each instant of time and which change and evolve in time. Among these entities there are the material (or
energetic) entities: these we imagine them to be present in space at any moment of time, as a kind of {\it
'substance'}. We shall refer to this hypothesis as to the {\it 'space contains reality hypothesis'}. This is an
important part of the intuitive view about the state of affairs around us. Within this intuitive view there are
many subtle questions that have occupied scientists and philosophers during the history of mankind. One of the
fundamental questions is that of the role of the observer in relation to this intuitive view on reality. We know
that all that we know about reality has come to us from our personal experience with this reality. It is also
clear that while we experience we also at the same time exert an influence, and sometimes we also create. Within
the intuitive view we also imagine reality to be independent of our experiencing it. To put it more directly,
reality would also exist if humankind would not be there, and if I would not be here now to experience it. Let us
call this belief the {\it 'realist hypothesis'}.  Newtonian mechanics and its elaborations had delivered at the
beginning of the foregoing century a complete theory of the inanimate world, wherein the role of the observer
literally could be neglected. The world presented itself as being a huge mechanic clockwork, evolving
deterministically according to the equations of Newton. We, the observers, did not have to be taken into account,
because the act of observation could be eliminated completely and hence did not have to be described in the
theory. This picture was also - independently of its realist aspects - in agreement with the intuitive view: it
was a fine and detailed mathematical modeling of this view and we shall refer to it as  {\it 'the classical
mechanical view'}. The {\it 'past-present-future hypothesis'} and the {\it 'space contains reality hypothesis'}
are satisfied in this {\it 'classical mechanical view'}. Indeed, reality is considered to be a collection of
material objects or substances, present at any moment of time at some place in a three dimensional Euclidean
space, and interacting with each other within this space, by means of interaction fields.

Within this Newtonian development many additional and fundamental aspects were added to the intuitive view. For
instance, classical mechanics is a deterministic theory: the state of the world at a certain moment, be it past,
present or future, is linked in a deterministic way to the state of the world at any earlier moment. Let us call
this {\it 'the determinist hypothesis'} and remark that no strict belief about determinism was originally
incorporated in the intuitive view.  The {\it 'classical mechanical view'} came into deep problems when Max
Planck made the first moves towards quantum mechanics. Quantum mechanics showed that the effect of the
measurement had to be taken into account in a crucial and non-reducible way for the description of the
micro-world; apparently, the old classical determinism was gone for good.

(1) Within the creation-discovery view and quantum mechanics, there are no reasons why the {\it
'past-present-future hypothesis'} should run into problems. Indeed, we can still consider reality as a process
that is ever changing and where the past is just the recollection of how this reality has been, and the future
the imagining of possible ways that this reality can become.

(2) There is also no problem with the {\it deterministic hypothesis}. There is no incompatibility at all between
quantum mechanics and a complete deterministic world as a whole, since the probabilities appearing in the quantum
theory can be explained as being due to a lack of knowledge about the interaction between the measuring apparatus
and the entity during the measurement, and hence are of epistemic nature. 

(3) The {\it 'space contains reality hypothesis'}, as we have explained in much detail, is the one that in our
opnion has to be abandoned. Reality is not contained within space. Space is a momentaneous cristalization of a
theatre for reality where the motions and interactions of the macroscopical material and energetic entities take
place. But other entities - like quantum entities for example - 'take place' outside space, or - and this would
be another way of saying the same thing - within a space that is not the three dimensional Euclidean space. 

(4) Quantum mechanics is not in contradiction with the {\it 'realist hypothesis'}. It is possible to believe that
reality exists independently of our observing and measuring it, and also that it would be there if there were no
humans to observe and influence it. But, as we have said already, this reality is is not contained within
Euclidian space. 

(5) When we consider relativity theory the situation is very subtle. For an ether interpretation of relativity
theory, there is no problem at all, since my personal reality remains identical with the three dimensional space
that is shared by all other humans, and hence can be considered as 'the present'. If an Einsteinian
interpretation of relativity theory is advocated, my personal reality has four dimensions, and the personal
realities of my fellow humans on earth also all have four dimensions. The {\it present-past-future hypothesis}
remains valid for all these personal realities, but it is not possible to fit them together into a single {\it
present-past-future scheme}. This shows that such a scheme is not without problems if we want to give a
description of the structure of reality that is not just the union of all the personal realities. We are at
present working hard to try and understand in which way these personal realities - all  four-dimensional and all
changing within a personal {\it past-present-future scheme} - can be fitted together inside a structure that would
account for a 'reality' which would be independent of all these personal realities.

\bigskip
\noindent
{\it Acknowledgments:} I want to thank my friend George Severne who read and commented parts  of
the article and with whom I also had many interesting discussions about the content.   

\section{References}

\smallskip
\noindent [1] {\bf Rauch H. et al.}, ``Test of a single crystal neutron
interferometer", {\it Phys. Lett.},{\bf 47 A}, 369, 1974.

\smallskip
\noindent [2] {\bf Rauch H. et al.}, ``Verification of coherent spinor
rotations of fermions", {\it Phys. Lett.},{\bf 54 A}, 425, 1975.

\smallskip
\noindent [3] {\bf Rauch H.}, ``Neutron interferometric tests of quantum
mechanics", {\it Helv. Phys. Acta}, {\bf 61}, 589, 1988.

\smallskip
\noindent [4]{\bf  Aerts D.}, ``An attempt to imagine parts of the reality
of the micro-world", in the proceedings of the conference {\it Problems in
Quantum Physics ; Gdansk '89,} World Scientific Publishing Company,
Singapore, 3, 1990.

\smallskip
\noindent [5] {\bf Aerts D. and Reignier J.}, ``The spin of a quantum
entity and problems of non-locality", in the proceedings of the {\it Symposium
on the Foundations of modern Physics 1990,} Joensuu, Finland, World Scientific
Publishing Company, Singapore, 9, 1990.

\smallskip
\noindent [6] {\bf Aerts D. and Reignier J.}, ``On the problem of
non-locality in quantum mechanics", {\it Helv. Phys. Acta}, {\bf 64}, 527, 1991.

\smallskip
\noindent [7] {\bf Aerts, D.}, ``A possible explanation for the
probabilities of quantum mechanics and a macroscopic situation that violates
Bell inequalities", in {\it Recent Developments in Quantum Logic,} eds. P.
Mittelstaedt et al., in Grundlagen der Exacten Naturwissenschaften, vol.6,
Wissenschaftverlag, Bibliographisches Institut, Mannheim, 235, 1983.

\smallskip
\noindent [8] {\bf Aerts, D.}, ``A possible explanation for the
probabilities of quantum mechanics", {\it J. Math. Phys.}, {\bf 27}, 202, 1986.

\smallskip
\par \noindent [9] {\bf Aerts, D.}, ``The origin of the
non-classical character of the quantum probability model", in {\it
Information, Complexity, and Control in Quantum Physics},  eds. Blanquiere, A., et al, Springer-Verlag, 1987.

\smallskip
\noindent [10] {\bf Aerts D.}, ``The construction of reality and its
influence on the  understanding of quantum structures", {\it Int. J. Theor. Phys.},
{\bf 31},  1815, 1992.

\smallskip
\par \noindent [11] {\bf Aerts, D.}, ``Quantum structures, separated
physical entities and probability",{\it Found. Phys.}, {\bf 24}, 1227, 1994.

\smallskip
\noindent [12] {\bf Aerts, D.}, Quantum structures: an attempt to
explain the origin of their appearance in  nature, {\it Int. J. Theor. Phys.} {\bf
34}, 1165, 1995.

\smallskip
\noindent [13] {\bf de Broglie L.}, ``Sur la possibilit\'e de relier les ph\'enom\`enes
d'interference et de diffraction \`a la th\'eorie des quanta de lumi\`ere", {\it Comptes Rendus},
{\bf 183}, 447, 1926.

\smallskip
\noindent [14] {\bf Bohm D. and Vigier J.P.}, ``Model of the causal interpretation of quantum
theory in terms of a fluid with irregular fluctuations", {\it Physical Review}, {\bf 96}, 208.

\smallskip
\noindent [15] {\bf Bohm D.},{\it Wholeness and the implicate order}, ARK Edition, 1983.

\smallskip
\noindent [16] {\bf Heisenberg W.}, ``Uber quantentheoretische Umdeuting kinematischer und
mechanischer Beziehungen", {\it Zeitsschrift f\"ur Physik}, {\bf 33}, 879, 1925.

\smallskip
\noindent [17] {\bf Schr\"odinger, E.}, ``Quantisiering als Eigenwertproblem", {\it Ann. der Phys.}, {\bf 79},
361, 1926.

\smallskip
\noindent [18] {\bf Dirac P.A.M.}, {\it The principles of quantum mechanics}, Oxford University
Press, 1930.

\smallskip
\noindent [19] {\bf Von Neumann J.}, {\it Mathematische Grundlagen der Quantenmechanik}, Berlin,
Springer, 1932.

\smallskip
\noindent [20] {\bf Jauch J.M.}, {\it Foundations of Quantum Mechanics}, Addison Wesley, 1968.

\smallskip
\noindent [21] {\bf Piron C.}, {\it Foundations of Quantum Physics}, Benjamin, 1976.

\smallskip
\noindent [22] {\bf Ludwig G.}, {\it Foundations of Quantum Mechanics}, (two volumes), Springer-Verlach, New
York, Berlin, 1983/5

\smallskip
\noindent [23] {\bf Randall C. and Foulis D.}, ``The 
operational approach to quantum
mechanics", in {\it Physical theory as logico-operational 
structure}, ed. Hooker, C. A.,
Reidel, 1979.

\smallskip
\noindent [24] {\bf Randall C. and Foulis D.}, ``A mathematical 
language for quantum physics", in {\it Les Fondements de la M\'ecanique 
Quantique}, eds. Gruber, C., {\it et al}, A.V.C.P., case postale 101, 1015 Lausanne, Suisse,  1983.

\smallskip
\noindent [25] {\bf Mittelsteadt P.}, {\it Quantum Logic}, Reidel, Dordrecht, 1978.

\smallskip
\noindent [26] {\bf Segal I.E.}, {\it Ann. Math.}, {\bf 48}, 930, 1947.

\smallskip
\noindent [27] {\bf Emch G.G.},{\it Mathematical and conceptual 
foundations of 20th century
physics,} North-Holland, Amsterdam, 1984.

\smallskip
\noindent [28] {\bf Feynman, R.P., Leighton, R.B. and Sands, M.}, {\it The Feynman Lectures on Physics, volume III},
Addison-Wesley Publishing Company, Reading, Massachusetts.

\smallskip
\noindent [29] {\bf Aerts D.}, {\it The one and the many}, Doctoral Thesis, Free University of
Brussels, Pleinlaan 2, 1050 Brussels, 1981.

\smallskip
\noindent [30] {\bf Aerts D.}, ``Description of many physical entities without the paradoxes
encountered in quantum mechanics", {\it Found. Phys.}, {\bf 12}, 1131, 1982.

\smallskip
\noindent [31] {\bf Aerts D.}, ``Classical theories and non classical theories as
a special case of a more general theory", {\it J. math. Phys.}, {\bf 24}, 2441, 1983.

\smallskip
\noindent
[32] {\bf Aerts, D.}, ``Foundations of physics: a general realistic and
operational approach", {\it Int. J. Theor. Phys.}, {\bf 38}, 289, 1999, and quant-ph/0105109.

\smallskip 
\noindent [33] {\bf Coecke B.}, ``Hidden measurement representation for quantum entities described by finite
dimensional complex Hilbert spaces", {\it Found. Phys.}, {\bf 25}, 1185, 1995.

\smallskip 
\noindent [34] {\bf Coecke B.}, ``Generalisation of the proof on the existence of hidden measurements to experiments with
an infinite set of outcomes", {\it Found. Phys. Lett.}, {\bf  8}, 437, 1995.

\smallskip 
\noindent [35] {\bf Coecke B.}, ``Hidden measurement model for pure and mixed states of quantum physics in Euclidean
space", {\it Int. J. Theor. Phys.}, {\bf 34}, 1313, 1995.

\smallskip
\noindent [36] {\bf Birkhoff G. and Von Neumann J.}, {\it J. Ann. Math.}, {\bf 37}, 823, 1936.

\smallskip
\noindent [37] {\bf Gudder S. P.}, {\it Quantum Probability}, Academic Press, Inc.
Harcourt Brave Jovanovitch, Publishers, 1988.

\noindent [38] {\bf Accardi  L.}, ``On the statistical meaning of the complex numbers in
quantum mechanics", {\it Nuovo Cimento}, {\bf 34}, 161, 1982. 

\smallskip
\noindent [39] {\bf Pitovski, I.}, {\it Quantum Probability - Quantum Logic}, Springer
Verlach, 1989.

\smallskip
\noindent [40] {\bf Aerts, D. and D'Hooghe B.}, ``Operator structure of a non-quantum and a non-classical
system", {\it Int. J. Theor. Phys.}, {\bf 35}, 2241, 1996.

\smallskip
\noindent [41] {\bf D'Hooghe B.}, ``From quantum to classical: a study of the effect of
varying fluctuations in the measurement context and state transitions due to
experiments", doctoral dissertation, Brussels Free University, 2000.

\smallskip
\par \noindent [42] {\bf Aerts D. and Van Bogaert B.}, ``A mechanical classical laboratory
situation with a quantum logic structure", {\it Int. J. Theor. Phys.}, {\bf 31}, 1839, 1992.

\smallskip
 \noindent [43] {\bf Aerts D., Durt T. and Van Bogaert B.}, ``A physical example of quantum fuzzy
sets, and the classical limit", {\it Tatra Montains Math. Publ.}, {\bf 1}, 5, 1992.

\smallskip
\noindent  [44] {\bf Aerts D., Durt T., Grib A. A., Van Bogaert B. and Zapatrin R. R.}, ``Quantum
structures in macroscopic reality", {\it Int. J. Theor. Phys.}, {\bf 32}, 489, 1993.

\smallskip
\par \noindent [45] {\bf Aerts D., Durt T. and Van Bogaert B.}, ``Quantum
probability, the classical limit and non-locality", in {\it Symposium on the Foundations of
Modern Physics}, ed. Hyvonen, T., World Scientific, Singapore, 1993.

\smallskip
\par \noindent [46] {\bf Aerts D. and Durt T.}, ``Quantum, classical and intermediate, an
illustrative example", {\it Found. Phys.}, {\bf 24}, 1994. 
\smallskip

\noindent [47] {\bf Aerts D. and Durt T.}, ``Quantum, classical and intermediate; a
measurement model", in the proceedings of the {\it International Symposium on the Foundations of Physics 1994,
Helsinki}, eds. Montonen, C., {\it et al}, Editions
Frontieres, Gives Sur Yvettes, France, 1994.

\smallskip
\par \noindent [48] {\bf Aerts D.}, ``The entity and modern physics: the
creation discovery view of reality", in {\it Interpreting Bodies:
Classical  and Quantum Objects in Modern Physics}, ed. Castellani, E.,
Princeton University Press, Princeton, 1998.

\smallskip
\par \noindent [49] {\bf Aerts  D. and Coecke  B.}, ``The creation-discovery view: towards a
possible explanation of quantum reality", in {\it Language,
Quantum, Music}, eds. M.L. Dalla Chiara, Kluwer Academic, Dordrecht, 1999.

\smallskip
\par \noindent  [50] {\bf Aerts D. and Aerts S.}, ``The hidden measurement
approach and conditional probabilities", in
{\it Fundamental Problems in Quantum Physics II}, eds. Ferrero, M. and van
der Merwe, A., Kluwer Academic, Dordrecht, 1996.

\smallskip
\par \noindent [51] {\bf Aerts, D.}, ``The hidden measurement formalism: what can be explained
and where paradoxes remain",  {\it Int. J.  Theor. Phys.}, {\bf 37}, 291, 1998, and quant-ph/0105126.

\smallskip
\noindent [52] {\bf Aerts D. and Aerts S.,} ``Applications of quantum statistics in
psychological studies of decision processes",  {\it Foundations of Science}, {\bf
1}, 85, 1994.

\smallskip
\noindent [53] {\bf Schr\"odinger E.}, ``Die gegenw Quantenmechanik", {\it Naturwissenschaften}, {\bf 23}, 807,
823 and 844, 1935.

\smallskip
\noindent [54] {\bf Valckenborgh F.}, ``Operational axiomatics and compound systems", in {\it Current Research in
Operational Quantum Logic}, eds. Coecke, B., Moore, D. and Wilce, A., Kluwer Academic, Dordrecht, 2000.

\smallskip
\noindent [55] {\bf Aerts D.}, ``How do we have to change quantum mechanics in order to describe
separated systems", in {\it The wave-particle Dualism}, eds. Diner, S., {\it et al}, D. Reidel
Publishing Company, Dordrecht, 419, 1984.

\smallskip
\noindent [56] {\bf Aerts D.}, ``The physical origin of the Einstein Podolsky Rosen paradox",
in {\it Open Questions in Quantum Physics,} eds. Tarozzi, G., and van der Merwe, A., D. Reidel
Publishing Company, Dordrecht, 33, 1985.

\smallskip
\noindent [57] {\bf Aerts D.}, ``The physical origin of the EPR paradox and how to violate Bell
inequalities by macroscopic systems", in {\it On the Foundations of modern Physics,}
eds. Lathi, P. and Mittelstaedt, P., World Scientific, Singapore, 305, 1985.

\smallskip
\par \noindent [58]  {\bf Aerts  D.,
Coecke B., Durt T. and Valckenborgh F.}, {\it Quantum,
classical and intermediate I : a model on the Poincar\'e sphere},
 Tatra Mt. Math. Publ., {\bf 10}, 225, 1997.

\smallskip
\par \noindent [58]  {\bf Aerts  D.,
Coecke B., Durt T. and Valckenborgh F.}, {``Quantum,
classical and intermediate I : a model on the Poincar\'e sphere",
 {\it Tatra Mt. Math. Publ.}, {\bf 10}, 225, 1997. 

\smallskip
\par \noindent [59]  {\bf Aerts  D., Coecke B., Durt T.,
and Valckenborgh F.}, ``Quantum,
classical and intermediate II : the vanishing vector space structure", 
{\it Tatra Mt. Math. Publ.}, {\bf 10}, 241, 1997.

\smallskip
\noindent [60] {\bf Aerts D., Aerts S., Durt T. and L\'ev\^eque O.}, ``Classical and quantum probability in the
$\epsilon$-model", {\it Int. J. Theor. Phys.}, {\bf 38}, 407, 1999.

\smallskip
\noindent [61] {\bf Einstein, A.}, Ann. Phys.,
{\bf 17}, 891,1905.

\smallskip
\noindent [62] {\bf Misner C. W., Thorne K. S. and  Wheeler J. A.}, {\it
Gravitation}, W. H. Freeman and Company, San Francisco, 1973.

\smallskip
\noindent [63] {\bf Aerts, D.}, ``Framework for possible unification of quantum and relativity theories",
{\it Int. J. Theor. Phys.}, {\bf 35}, 2431, 1996.

\smallskip
\noindent [64]{\bf  Aerts, D.}, ``Relativity theory: what is reality?", {\it Found. Phys.}, {\bf 26}, 1627, 1996.

\end{document}